\documentstyle[sprocl,epsfig]{article}
\makeatletter
\def\@cite#1#2{{[{#1}]\if@tempswa\typeout
{IJCGA warning: optional citation argument
ignored: `#2'} \fi}}

\newcount\@tempcntc
\def\@citex[#1]#2{\if@filesw\immediate\write\@auxout{\string\citation{#2}}\fi
  \@tempcnta\z@\@tempcntb\m@ne\def\@citea{}\@cite{\@for\@citeb:=#2\do
    {\@ifundefined
       {b@\@citeb}{\@citeo\@tempcntb\m@ne\@citea\def\@citea{,}{\bf ?}\@warning
       {Citation `\@citeb' on page \thepage \space undefined}}%
    {\setbox\z@\hbox{\global\@tempcntc0\csname b@\@citeb\endcsname\relax}%
     \ifnum\@tempcntc=\z@ \@citeo\@tempcntb\m@ne
       \@citea\def\@citea{,}\hbox{\csname b@\@citeb\endcsname}%
     \else
      \advance\@tempcntb\@ne
      \ifnum\@tempcntb=\@tempcntc
      \else\advance\@tempcntb\m@ne\@citeo
      \@tempcnta\@tempcntc\@tempcntb\@tempcntc\fi\fi}}\@citeo}{#1}}
\def\@citeo{\ifnum\@tempcnta>\@tempcntb\else\@citea\def\@citea{,}%
  \ifnum\@tempcnta=\@tempcntb\the\@tempcnta\else
   {\advance\@tempcnta\@ne\ifnum\@tempcnta=\@tempcntb \else \def\@citea{--}\fi
    \advance\@tempcnta\m@ne\the\@tempcnta\@citea\the\@tempcntb}\fi\fi}
\makeatother

\def\lsim{\mathrel{\raise.3ex\hbox{$<$\kern-.75em\lower1ex\hbox{$\sim$}}}}
\def\gsim{\mathrel{\raise.3ex\hbox{$>$\kern-.75em\lower1ex\hbox{$\sim$}}}}
\def\bold#1{\setbox0=\hbox{$#1$}%
     \kern-.025em\copy0\kern-\wd0
     \kern.05em\copy0\kern-\wd0
     \kern-.025em\raise.0433em\box0 }

\def\ifmath#1{\relax\ifmmode #1\else $#1$\fi}
\def\half{\ifmath{{\textstyle{1 \over 2}}}}
\def\threehalf{\ifmath{{\textstyle{3 \over 2}}}}

\def\third{\ifmath{{\textstyle{1 \over 3}}}}

\def\fivethirds{{\textstyle{5 \over 3}}}

\def\nmess{N_m}
\def\vev#1{\langle #1 \rangle}
\def\lam{\lambda}

\def\Eq#1{Eq.~(\ref{#1})}

\def\bml{\hbox{$B\!\!-\!\!L$}}
\def\gtino{\wt g_{3/2}}
\def\mgtino{m_{\gtino}}
\def\mplanck{M_{\rm P}}
\def\mpl{\mplanck}
\def\sur{{\wt u_R}}
\def\msur{{m_{\sur}}}

\def\sbl{{\wt b_L}}
\def\sbr{{\wt b_R}}
\def\msbl{m_{\sbl}}
\def\msbr{m_{\sbr}}
\def\sq{\wt q}

\def\msq{m_{\sq}}
\def\slep{\wt \ell}

\def\mslep{m_{\slep}}
\def\slepl{\wt \ell_L}
\def\mslepl{m_{\slepl}}
\def\slepr{\wt \ell_R}
\def\mslepr{m_{\slepr}}

\def\sel{\wt e}

\def\sell{\wt e_L}
\def\msell{m_{\sell}}
\def\selr{\wt e_R}
\def\mselr{m_{\selr}}

\def\cpmtwo{\wt \chi^{\pm}_2}

\def\mcpmtwo{m_{\cpmtwo}}

\def\mth{m_{3/2}}
\def\delgs{\delta_{GS}}
\def\kpr{K^\prime}

\def\Twoloop{Two-loop/RGE-improved}

\def\mhi{m_{h_1^0}}
\def\etmiss{/ \hskip-8pt E_T}
\def\etmin{/ \hskip-8pt E_T^{\rm min}}

\def\mhalf{m_{1/2}}
\def\gl{\wt g}
\def\mgl{m_{\gl}}

\def\etc{{\it etc.}}

\def\h{h}
\def\a{a}
\def\mh{m_{\h}}
\def\ma{m_{\a}}
\def\gamh{\Gamma_{\h}^{\rm tot}}

\def\etc{{\em etc.}}

\def\eg{{\it e.g.}}
\def\etal{{\it et al.}}
\def\mhalf{m_{1/2}}

\def\stop{\wt t}
\def\stopone{\wt t_1}
\def\stoptwo{\wt t_2}
\def\mstop{m_{\stop}}

\def\mstopone{m_{\stopone}}
\def\mstoptwo{m_{\stoptwo}}

\def\sbot{\wt b}

\def\sbl{{\wt b_L}}
\def\sbr{{\wt b_R}}
\def\msbl{m_{\sbl}}
\def\msbr{m_{\sbr}}

\def\slep{\wt \ell}

\def\mslep{m_{\slep}}
\def\slepl{\wt \ell_L}
\def\mslepl{m_{\slepl}}
\def\slepr{\wt \ell_R}
\def\mslepr{m_{\slepr}}

\def\msusy{m_{\rm SUSY}}

\def\susy{{\rm SUSY}}

\def\etc{{\it etc.}}

\def\mrzero{m_{R^0}}
\def\etc{{\em etc.}}

\def\eg{{\it e.g.}}
\def\etal{{\it et al.}}
\def\mhalf{m_{1/2}}
\def\gl{\wt g}
\def\mgl{m_{\gl}}
\def\stop{\wt t}
\def\stopone{\wt t_1}
\def\mstop{m_{\stop}}

\def\mstopone{m_{\stopone}}

\def\slep{\wt \ell}

\def\mslep{m_{\slep}}
\def\slepl{\wt \ell_L}
\def\mslepl{m_{\slepl}}
\def\slepr{\wt \ell_R}
\def\mslepr{m_{\slepr}}
\def\sbot{\wt b}

\def\hsm{h_{\rm SM}}
\def\mhsm{m_{\hsm}}
\def\hl{h^0}
\def\hh{H^0}
\def\ha{A^0}
\def\hp{H^+}
\def\hm{H^-}
\def\hpm{H^{\pm}}
\def\mhl{m_{\hl}}
\def\mhh{m_{\hh}}
\def\mha{m_{\ha}}

\def\mhpm{m_{\hpm}}
\def\tanb{\tan\beta}

\def\mt{m_t}
\def\mb{m_b}
\def\mz{m_Z}
\def\mw{m_W}
\def\mgut{M_U}
\def\mx{M_X}

\def\wpm{W^{\pm}}

\def\chitil{\wt\chi}
\def\cnone{\wt\chi^0_1}
\def\cnonestar{\wt\chi_1^{0\star}}
\def\cntwo{\wt\chi^0_2}
\def\cnthree{\wt\chi^0_3}

\def\snu{\wt\nu}
\def\snul{\wt\nu_L}

\def\msnu{m_{\snu}}
\def\mcnone{m_{\cnone}}
\def\mcntwo{m_{\cntwo}}
\def\mcnthree{m_{\cnthree}}

\def\h{h}
\def\mh{m_{\h}}
\def\wt{\widetilde}
\def\wh{\widehat}
\def\cpone{\wt \chi^+_1}
\def\cmone{\wt \chi^-_1}
\def\cpmone{\wt \chi^{\pm}_1}
\def\mcpone{m_{\cpone}}
\def\mcpmone{m_{\cpmone}}

\def\cpmtwo{\wt \chi^{\pm}_2}

\def\mcpmtwo{m_{\cpmtwo}}

\def\staur{\wt \tau_R}
\def\staul{\wt \tau_L}
\def\stau{\wt \tau}
\def\mstaur{m_{\staur}}

\def\MPL #1 #2 #3 {{\sl Mod.~Phys.~Lett.}~{\bf#1} (#3) #2}
\def\NPB #1 #2 #3 {{\sl Nucl.~Phys.}~{\bf B#1} (#3) #2}
\def\PLB #1 #2 #3 {{\sl Phys.~Lett.}~{\bf B#1} (#3) #2}
\def\PR #1 #2 #3 {{\sl Phys.~Rep.}~{\bf#1} (#3) #2}
\def\PRD #1 #2 #3 {{\sl Phys.~Rev.}~{\bf D#1} (#3) #2}
\def\PRL #1 #2 #3 {{\sl Phys.~Rev.~Lett.}~{\bf#1} (#3) #2}
\def\RMP #1 #2 #3 {{\sl Rev.~Mod.~Phys.}~{\bf#1} (#3) #2}
\def\ZPC #1 #2 #3 {{\sl Z.~Phys.}~{\bf C#1} (#3) #2}
\def\IJMP #1 #2 #3 {{\sl Int.~J.~Mod.~Phys.}~{\bf#1} (#3) #2}

\def\wpm{W^{\pm}}
\def\hpm{H^{\pm}}

\def\call{{\cal L}}
\def\wtil{\widetilde}

\def\tauptaum{\tau^+\tau^-}

\def\lam{\lambda}
\def\br{BF}
\def\tauptaum{\tau^+\tau^-}

\def\gam{\gamma}
\def\sigrts{\sigma_{\tiny\rts}^{}}

\def\etal{{\it et al.}}
\def\etc{{\it etc.}}

\def\anti{\overline}
\def\epem{e^+e^-}
\def\zstar{Z^\star}
\def\wstar{W^\star}

\def\mupmum{\mu^+\mu^-}

\def\rts{\sqrt s}
\def\ie{{\it i.e.}}
\def\eg{{\it e.g.}}

\def\anti{\overline}

\def\mw{m_W}
\def\mz{m_Z}
\def\h{h}
\def\mh{m_{\h}}
\def\gamh{\Gamma_{\h}^{\rm tot}}

\def\a{a}
\def\ma{m_{\a}}
\def\hsm{h_{SM}}
\def\mhsm{m_{\hsm}}
\def\gamhsm{\Gamma_{\hsm}^{\rm tot}}
\def\tanb{\tan\beta}
\def\hl{h^0}
\def\mhl{m_{\hl}}
\def\gamhl{\Gamma_{\hl}^{\rm tot}}
\def\ha{A^0}
\def\mha{m_{\ha}}

\def\hh{H^0}
\def\mhh{m_{\hh}}

\def\fbi{~{\rm fb}^{-1}}
\def\fb{~{\rm fb}}
\def\pbi{~{\rm pb}^{-1}}

\def\mev{~{\rm MeV}}
\def\gev{~{\rm GeV}}
\def\tev{~{\rm TeV}}
\def\stop{\widetilde t}
\def\mstop{m_{\stop}}
\def\mt{m_t}
\def\mb{m_b}

\def\hi{\h_1}

\def\mhi{m_{\hi}}

\newcommand{\nc}{\newcommand}
\nc{\beq}{\begin{equation}}   \nc{\eeq}{\end{equation}}
\nc{\bea}{\begin{eqnarray}}   \nc{\eea}{\end{eqnarray}}
\nc{\baa}{\begin{array}}      \nc{\eaa}{\end{array}}
\nc{\bit}{\begin{itemize}}    \nc{\eit}{\end{itemize}}
\nc{\ben}{\begin{enumerate}}  \nc{\een}{\end{enumerate}}
\nc{\bce}{\begin{center}}     \nc{\ece}{\end{center}}
\def\beqa{\begin{eqnarray}}
\def\eeqa{\end{eqnarray}}

\def\Twoloop{Two-loop/RGE-improved}

\def\tanb{\tan\beta}

\begin{document}
\font\fortssbx=cmssbx10 scaled \magstep2
\hbox to \hsize{
$\vcenter{
\hbox{\fortssbx University of California - Davis}
}$
\hfill
$\vcenter{
\hbox{\bf UCD-98-2} 
\hbox{\bf hep-ph/9801417}
\hbox{January, 1998}
}$
}
\medskip
\pagestyle{empty}
\title{SEARCHING FOR LOW-ENERGY SUPERSYMMETRY~\footnote{To appear in 
{\it Quantum Effects in the MSSM}, the Proceedings of the ``International
Workshop on Quantum Effects in the MSSM'', 
UAB, Barcelona, September 9--13, 1997,
edited by J. Sola (World Scientific Publishing).}
}
\author{ John F. Gunion}
\address{Davis Institute for High Energy Physics,   \\
Department of Physics, \\  University of California, Davis, CA 95616}
\maketitle\abstracts{A review of supersymmetry theory and
phenomenology is presented. Topics discussed include:
gravity-mediated (SUGRA) and
gauge-mediated (GMSB) supersymmetry breaking models;
an overview of non-universal soft-supersymmetry-breaking
masses and the resulting experimental implications; the phenomenology of and
constraints on the possibility that a massive gluino is the
lightest supersymmetric particle; current status of the phenomenology
of the Higgs bosons of the minimal and next-to-minimal supersymmetric models; 
and, the signals for GMSB and/or
R-parity-violating models coming from delayed decays of the 
lightest supersymmetric partner of the Standard Model particles.}

\section{Introduction}

There are good reasons why
supersymmetry (SUSY) is regarded by many as the most
attractive candidate for the next level of
physics beyond the Standard Model (SM).
First, there are important aesthetic motivations for \susy.
i) It is the only non-trivial extension of the Lorentz group at the 
heart of quantum field theory.
ii) If \susy\ is formulated as a {\it local} symmetry, it 
becomes a super-gravity (SUGRA) theory 
that reduces to general relativity in the appropriate limit.
iii) \susy\ appears in Superstrings.
Second, although \susy\ must be a broken
symmetry since we have not yet observed supersymmetric particles,
there are numerous indications that it will be realized at
energies $\lsim {\rm few}~\tev$.
i) String theory solutions with a non-supersymmetric {\it ground} state are
quite problematic.
ii)
\susy\ solves the hierarchy problem, 
\eg\ $m_{\rm Higgs}^2<(1\tev)^2$ (as required
to avoid a non-perturbative $WW$ sector) 
via spin-1/2 loop cancellation of spin-0
loop quadratic divergences {\it provided} the masses 
(generically denoted by $\msusy$) of
supersymmetric partners are $\lsim {\rm few}~\tev$.
iii) The minimal supersymmetric model (MSSM)
leads to gauge coupling unification (at $\mgut\sim {\rm few}\times
10^{16}\gev$) if $\msusy\lsim 1-10\tev$.
iv) In the MSSM, electroweak symmetry breaking (EWSB) 
can be a natural result of renormalization group evolution (RGE)
if $\mt\sim 175\gev$.
v) In many cases, the lightest supersymmetric particle
is a natural cold dark matter candidate.

However, there are many phenomenologically viable supersymmetry models.
These and their experimental implications will be the primary
subject of these lectures. The present review is an expanded version
of Ref.~\cite{gunhabper}. Full referencing for 
the more introductory material can be found there. Here, I will include
detailed referencing only for some of the relatively recent developments and
for the collider phenomenology discussions.

\section{The Minimal Supersymmetric Model}

The minimal supersymmetric extension of the Standard Model 
consists of taking the Standard Model
and adding the corresponding supersymmetric partners.
The SM particles and corresponding sparticle mass eigenstates are listed below.
\begin{eqnarray}
\left[u,d,c,s,t,b\right]_{\phantom{L,R}} & ~~~~g~~~~ & \underbrace{\wpm,\hpm}
~~~\underbrace{\gam,Z,H_1^0,H_2^0}\nonumber \\
\left[{\tilde u},{\tilde d},{\tilde c},{\tilde s},{\tilde t},{\tilde
b}\right]_{L,R}
 & ~~~~{\tilde g}~~~~& ~~~~\chitil_{1,2}^{\pm}~~~~~~~~~~\chitil_{1,2,3,4}^0
\nonumber
\end{eqnarray}
In addition, the MSSM contains two hypercharge $Y=\pm 1$ Higgs
doublets, which is the minimal structure for the Higgs sector of an
anomaly-free supersymmetric extension of the Standard Model.
The supersymmetric structure of the theory also requires (at least) two
Higgs doublets to generate mass for both ``up''-type and ``down''-type
quarks (and charged leptons). The physical Higgs mass eigenstates
then comprise two CP-even Higgs
bosons, $\hl,\hh$ (with $\mhl\leq\mhh$), a CP-odd Higgs boson, $\ha$,
and a charged Higgs pair, $\hpm$.

In the MSSM, all renormalizable supersymmetric interactions
consistent with (global) \bml\
conservation are included.  They are parameterized by:
\begin{itemize}
\item
 SU(3)$\times$SU(2)$\times$U(1) gauge couplings, $g_3$, $g_2\equiv g$,
and $g_1\equiv\sqrt{\fivethirds}\,g^\prime$;
\item
(complex) Higgs-quark Yukawa coupling matrices, ${\bf y_u}$,
${\bf y_d}$, and ${\bf y_e}$; and
\item
a supersymmetric Higgs mass parameter, $\mu$.
\end{itemize}
Since the MSSM is a model of three generations of quarks, leptons and
their superpartners, ${\bf y_u}$, ${\bf y_d}$, and ${\bf y_e}$ are
complex $3\times 3$ matrices.  
However, as discussed below, not all these degrees of
freedom are physical.

As a consequence of \bml\ invariance, the MSSM possesses
a multiplicative R-parity invariance, such that
all the ordinary Standard Model
particles have even R-parity, whereas the corresponding
supersymmetric partners have odd R-parity.

In the MSSM, supersymmetry breaking is accomplished by including the
most general renormalizable soft-supersymmetry-breaking terms
consistent with the SU(3)$\times$SU(2)$\times$U(1) gauge symmetry and
R-parity conservation. These terms
parameterize our ignorance of the fundamental mechanism of
super\-symmetry breaking.  It is here
where most of the new supersymmetric model parameters reside.
As already noted, if super\-symmetry is relevant
for explaining the scale of electroweak interactions, then the
mass parameters introduced by the soft-super\-symmetry-breaking
terms should be in the TeV range or below.

The soft-supersymmetry-breaking terms of the MSSM include:
\begin{itemize}
\item
gaugino Majorana masses $M_3$, $M_2$, and $M_1$,
associated with the SU(3), SU(2), and U(1) subgroups of the Standard Model;
\item
diagonal Higgs squared-mass terms, $m_{H_u}^2$, $m_{H_d}^2$;
\item
an off-diagonal Higgs squared-mass term, $b$ (the ``B''-term);
\item
(hermitian) squark and slepton squared-mass matrices,
${\bf m_{Q}^2}$, ${\bf m_{\anti u}^2}$,
${\bf m_{\anti d}^2}$,
${\bf m_{L}^2}$, ${\bf m_{\anti e}^2}$;
\item
(complex) matrix $A$-parameters,
${\bf a_u}$, ${\bf a_d}$, ${\bf a_e}$,
parameterizing Higgs-squark-squark and Higgs-slepton-slepton
trilinear interactions.
\end{itemize}
As above, the quantities specified by boldface letters are $3\times 3$
matrices (in generation space).
Note that after minimizing the Higgs potential, one can trade in the
three Higgs parameters $m_{H_u}^2$, $m_{H_d}^2$ and $b$
for two neutral Higgs field vacuum expectation values $v_u$ and
$v_d$ and one of the physical Higgs masses.
Equivalently, one fixes $v^2\equiv v_u^2+v_d^2=4\mw^2/g^2
=(246~{\rm GeV})^2$ and introduces the parameter $\tanb\equiv v_u/v_d$.

There are many unphysical degrees of freedom among 
the MSSM parameters listed above. These can be eliminated by
expressing interaction eigenstates in terms of the mass eigenstates,
with an appropriate redefinition of the MSSM fields to remove unphysical
phases.  In the end, the MSSM
possesses 124 truly independent parameters.  Of these, 18 parameters
correspond to Standard Model parameters
(including the QCD vacuum angle $\theta_{\rm QCD}$), one corresponds to
one of the Higgs sector masses (the analogue of the Standard Model
Higgs mass), and 105 are genuinely new parameters of the model.
These include:
five real parameters and three CP-violating phases in
the gauge/gaugino/Higgs/higgsino sector, 21 squark and slepton masses,
36 new real mixing angles to define the
squark and slepton mass eigenstates and 40 new CP-violating phases that
can appear in squark and slepton interactions.
The most general R-parity conserving minimal supersymmetric
extension of the Standard Model (without additional theoretical
assumptions) will be denoted henceforth as MSSM-124.

MSSM-124 is phenomenologically viable only for very special 
domains in the full parameter space such that the following
possible problems are absent:
(i) non-conservation of the separate lep\-ton numbers
L$_e$, L$_\mu$, and L$_\tau$; (ii) unsuppressed
flavor-changing neutral currents (FCNC's); and (iii)
new sources of CP-violation that are
inconsistent with the experimental bounds.  

There are a number of approaches for reducing the parameter freedom of
MSSM-124 in such a way as to avoid these problems. There are two low-energy
approaches.
First, one can assume that the squark and slepton squared-mass matrices
and the matrix $A$-parameters
are proportional to the $3\times 3$ unit matrix
(horizontal universality).
Alternatively, one can simply
require that all the aforementioned matrices are flavor-diagonal in a
basis where the quark and lepton mass matrices are diagonal (flavor
alignment).  In either case, L$_e$, L$_\mu$ and
L$_\tau$ are separately
conserved, while tree-level FCNC's are automatically absent.
In both cases, the number of free parameters characterizing the MSSM
is substantially less than 124.  Both
scenarios are phenomenologically viable, but there is no strong
theoretical basis for either scenario.

In high-energy approaches, one
treats the parameters of the MSSM as running parameters and imposes a
particular structure on the soft-super\-symmetry-breaking terms at
a common high-energy scale [such as the GUT scale $\mgut$ or the
reduced Planck scale $\mplanck=(8\pi G_N)^{-1/2}\sim 2.4\times 10^{18}\gev$].
Using the renormalization group equations, one can then
derive the low-energy MSSM parameters.
The initial conditions (at the appropriate high-energy scale)
for the renormalization group equations
depend on the mechanism by which supersymmetry
breaking is communicated to the effective low-energy theory.
One bonus of such an approach is that one of the diagonal Higgs
squared-mass
parameters is typically driven negative by renormalization group
evolution.  Thus, electroweak symmetry breaking is generated
radiatively, and the resulting electroweak symmetry-breaking scale is
intimately tied to the scale of low-energy supersymmetry breaking.

In the context of the MSSM, the high-energy approach is especially
well-motivated given the fact that the MSSM particle content
yields relatively exact gauge coupling unification (for
sparticle masses $\lsim 1-10\tev$). If the gauge couplings
become universal at $\mgut$, then it is quite natural to suppose
that the soft supersymmetry breaking parameters could also take
a simple form at $\mgut$. A number of points
concerning gauge coupling unification should be emphasized.
a) Perturbative unification requires that
the number of families in the $\lsim 1\tev$ mass range be $\leq 4$. 
b) Unification occurs only if there are exactly 2 Higgs doublets, as
in the MSSM, and possibly additional singlets,
below $\mgut$. Coupling unification in the presence of
additional doublets, triplets and/or higher Higgs representations
is only possible if appropriate 
additional matter fields with mass above $1-10\tev$
and below $\mgut$ are introduced.
c) The currently preferred smaller $\alpha_s(\mz)$ values are
most consistent with coupling unification if the `average' sparticle
mass $\msusy$ is above $1\tev$.
d) The fact that the predicted MSSM value of 
$\mgut$ is substantially below $\mplanck$
may prove difficult to accommodate in specific string models.

To summarize, 
the word ``minimal'' in MSSM refers to the minimal particle spectrum and
the associated R-parity invariance.  The MSSM
particle content must be supplemented by assumptions about the origin of
supersymmetry-breaking that lie outside the low-energy domain of the
model.  The corresponding parameter freedom is bad
if you want to know ahead of time exactly how to discover supersymmetry.
Machine builders and experimentalists must be prepared for a wide range of
possibilities as to how to see and fully explore SUSY.
But, it is good in the sense that once the 105 new parameters of the MSSM
have been experimentally determined 
you will have learned a great deal about the model.
For example, in models where supersymmetry-breaking is mediated
by gravitational interactions one can hope
to discover an underlying organization for the 105 parameters
(by evolving the low-$E$ parameters up to $\mgut$ or $\mplanck$) 
that will define the GUT/String model visible sector properties.
Different models lead to vastly different low-$E$ parameters and expectations.
The rest of the talk reviews some of the many possibilities
discussed to date.

\section{Constrained MSSM's: mSUGRA and GMSB}

In these theories, one posits a theory of dynamical
supersymmetry-breaking (DSB), where the DSB occurs in a sector that is
distinct from the fields of the MSSM.  The supersymmetry-breaking
inherent in the DSB sector is subsequently transmitted to the
MSSM spectrum by some mechanism.

Two theoretical scenarios have been examined in detail: gravity-mediated
and gauge-mediated supersymmetry breaking. In these approaches,
supersymmetry is spontaneously broken, in which case a massless Goldstone
fermion, the {\it goldstino}, arises.  Its coupling to a particle and
its superpartner is fixed by the supersymmetric Goldberger-Trieman relation:
\beq
\call_{\rm int}=-{1\over F}\,j^{\mu\alpha}\partial_\mu \wt G_\alpha+{\rm
h.c.}\,,
\label{gtrel}
\eeq
where $j^{\mu\alpha}$ is the supercurrent, which depends bilinearly on
all the fermion--boson superpartner pairs of the theory and
$\wt G_\alpha$ is the spin-1/2 goldstino field (with spinor index
$\alpha$).
$\sqrt{F}$ is the scale at which supersymmetry-breaking occurs
in the hidden sector (typically, $\sqrt{F}\gg\mz$).
When gravitational effects are included, the goldstino is ``absorbed''
by the {\it gravitino} ($\gtino $),
the spin-3/2 partner of the graviton.
By this super-Higgs mechanism, the goldstino is removed from the
physical spectrum and the gravitino acquires a mass ($m_{3/2}$).
In models where the gravitino mass is generated at tree-level one finds
\begin{equation}
m_{3/2}={F\over \sqrt 3\mpl}\,.
\label{mgravitino}
\end{equation}
The helicity $\pm\half$ components of
the gravitino behave approximately like the goldstino, whose
couplings to particles and their superpartners are determined by
\Eq{gtrel}.  In contrast, the helicity $\pm\threehalf$ components of
the gravitino couple with gravitational strength to particles and
their superpartners, and thus can be neglected.

In many models, the DSB sector is comprised of fields that
are completely neutral with respect to the Standard Model gauge group.
In such cases, the DSB sector is also called the ``hidden sector''.
The fields of the MSSM are said to reside in
the ``visible sector'', and the model is constructed such
that no renormalizable tree-level interactions exist between
fields of the visible and hidden sectors.  A third sector, the
so-called ``messenger sector'', is often employed in models to transmit
the supersymmetry-breaking from the hidden sector to the visible sector.
However, it is also possible to construct models in which the DSB sector
is not strictly hidden and contains fields that are charged with respect
to the Standard Model gauge group.

\subsection{Gravity-mediated supersymmetry breaking}

All particles feel the gravitational force.  In particular,
particles of the
hidden sector and the visible sector can interact via the exchange of
gravitons.  Thus, supergravity (SUGRA) models provide a natural mechanism
for transmitting the supersymmetry-breaking of the hidden sector to the
particle spectrum of the MSSM. In models of {\it gravity-mediated}
super\-symmetry breaking, gravity is the messenger of
supersymmetry-breaking.
The resulting low-energy
effective theory below the Planck scale consists of the unbroken MSSM
plus all possible soft-supersymmetry-breaking terms.

For example, the gaugino masses could arise primarily from
an effective Lagrangian term of the form
\begin{equation}
\call= \int d^2\theta W_a W_b {\wh \Phi_{ab}\over \mplanck} + h.c.
\sim  {\langle F_{\Phi} \rangle_{ab}\over \mplanck}\lam_a\lam_b\,,
\label{callform}
\end{equation}
where the $\lam_{a,b}$ ($a,b=1,2,3$) are the gaugino fields and $F_{\Phi}$
is the auxiliary component of a hidden sector superfield $\wh \Phi$
that appears linearly in the gauge kinetic function.
The scale $\mplanck$
appears in the denominator when communication between the hidden sector
and the MSSM sector is via gravitational interactions. 
$F$ in the previous section should be identified with $\vev{F_{\Phi}}$.
From \Eq{callform}\ we see that in order for
gauginos to have mass ${\cal O}(\mw)$, $\sqrt F\sim
\sqrt{\mw\mplanck}\sim 10^{10}\gev$ is required.
Such a large $\sqrt F$ value implies that 
the gravitino mass, \Eq{mgravitino}, is larger than the mass of the lightest
neutralino, $\cnone$. Since the gravitino is also {\it very} weakly interacting
when $\sqrt F$ is large, \Eq{gtrel}, 
it becomes phenomenologically irrelevant.

In a {\it minimal} supergravity (mSUGRA) framework, the soft-supersymmetry
breaking parameters at the Planck or GUT scale take a particularly
simple form. First, the gaugino mass parameters 
as well as the gauge couplings are presumed to unify at 
some scale $\mx$ ($=\mgut$ or $\mplanck$?):
\beq
\label{gunif}
M_1(\mx)=M_2(\mx)=M_3(\mx)\equiv m_{1/2}\,.
\eeq         
[For example, in the context of \Eq{callform}, the 
field $\wh \Phi$ of the hidden sector
would be presumed to be a flavor singlet, implying
$\vev{ F_{\Phi}}\propto c\delta_{ab}$.]
\Eq{gunif}\ implies that the
low-energy gaugino mass parameters satisfy:
\beq \label{gauginomassrelation}
 M_3 = {g^2_3\over g_2^2} M_2\simeq 3.5M_2,
\qquad M_1 = \fivethirds\tan^2\theta_W M_2\simeq 0.5M_2\,.
\eeq
Second, the scalar squared masses
and the $A$-parameters are taken to be flavor-diagonal and universal:
\beqa \label{plancksqmasses}
 &&{\bf m^2_{Q}} (\mpl) = {\bf m^2_{\anti u}}(\mpl) =
{\bf m^2_{\anti d}}(\mpl) = {\bf m^2_{L}}(\mpl) = {\bf m^2_{\anti e}}(\mpl)
=  m_0^2 {\bf 1}
\,,\nonumber\\
 &&m^2_{H_u}(\mpl) = m^2_{H_d}(\mpl) = m_0^2 \,,\nonumber \\
 && {\bf a_f}(\mpl) =A_0 {\bf y_f}(\mpl)\,,\qquad {\bf f}={\bf u,d,e}\,,
\eeqa
where ${\bf 1}$ is the $3\times 3$ identity matrix in generation space, and
the ${\bf y_f}$ are the Higgs-fermion Yukawa coupling matrices.
Finally, $\mu$, $b$ and the gaugino mass parameters
are assumed (rather arbitrarily) to be initially real at the high scale.

It is easy to count
the number of free parameters of mSUGRA. The low-energy values of
the MSSM parameters are determined by the MSSM renormalization group
equations and
the above initial conditions [Eqs. (\ref{gunif}) and (\ref{plancksqmasses})]. 
Thus, the mSUGRA model is fixed by the following
parameters: 18 Standard Model parameters (excluding the
Higgs mass), $m_0$, $m_{1/2}$, $A_0$, $\tanb$, and ${\rm sgn}(\mu)$ for
a total of 23 parameters.  This result was obtained by trading in the
parameters $b$ and $\mu^2$ for $v^2$
(which is counted as one of the 18 Standard Model
parameters) and $\tanb$. In this procedure the sign of
$\mu$ is not fixed, and so it remains an independent degree of freedom.
The requirement of radiative electroweak symmetry breaking
imposes an additional constraint on the possible range of the mSUGRA
parameters.  In particular, one finds that $1\lsim\tanb\lsim m_t/m_b$.
In principle, the mass of the gravitino,
$m_{3/2}$ (or equivalently, $\sqrt{F}$), is an additional independent
parameter. But, as discussed above,
the gravitino does not play any role in mSUGRA phenomenology.

Clearly, mSUGRA is much more predictive than MSSM-124.
In particular, one has only four genuinely
new parameters beyond the Standard Model
(plus a two-fold ambiguity corresponding to the sign of
$\mu$), from which one can predict the entire MSSM spectrum and its
interactions. It will be convenient to reference a number of
specific mSUGRA parameter sets. The first is
the `Standard' Snowmass96 Point, defined by
$m_0=200\gev$, $\mhalf=100\gev$, $A_0=0$, $\tanb=2$, $\mu<0$.
Other interesting parameter sets include
the String/SUGRA motivated
No-Scale ($\mhalf\neq0$, $m_0=A_0=0$) 
and Dilaton or Dilaton-Like ($\mhalf=-A_0=\sqrt 3 m_0$) models.
The latter are perhaps particularly worthy of considering since
to deviate significantly from
dilaton-like boundary conditions in Calabi-Yau and Orbifold models
requires that supersymmetry breaking be overwhelmingly
dominated by moduli fields other than the dilaton 
($\sin\theta_{\rm goldstino}\ll 1$). This is because
in the absence of loop corrections the dilaton boundary conditions
apply with $\mhalf=m_{3/2}\sin\theta_{\rm goldstino}$ and
it is only when $m_{3/2}\sin\theta_{\rm goldstino}$ is small compared
to loop corrections that we can move away from dilaton boundary conditions.
Vacuum stability against color breaking, might be a problem for
this model, but it currently appears that
metastability and/or non-perturbative effects can cure the apparent
difficulty. In both the no-scale and dilaton cases, two parameters
(usually chosen to be $\mhalf$ and $\tanb$) and the sign of $\mu$
are sufficient to completely specify the model.

The mSUGRA initial conditions of \Eq{plancksqmasses}\
correspond to a minimal SUGRA framework (specifically,
the kinetic energy terms for the gauge and matter fields are assumed
to take a minimal canonical form). Since there is no theoretical principle
that enforces such a minimal structure, 
perturbations of the mSUGRA initial
conditions consistent with phenomenology
({\it e.g}, avoiding the generation of dangerous FCNC's)
have been considered \cite{cmssmpert,snowtheory2}. For example, one can
introduce separate mass scales
for the Higgs and squark/slepton soft-supersymmetry-breaking masses.
One can also introduce non-universal scalar masses, but
restrict the size
of the non-universal terms to be consistent with phenomenology.
Finally, gaugino mass unification is {\it not} 
required by either theory or phenomenological constraints.
Specific examples of all these deviations arise in string-inspired models
\cite{nonuniv}.
Models which violate the mSUGRA initial conditions that 
appear to be of particular interest include:
\begin{itemize}
\item Strange-event-motivated models.
For example, the phenomenology of $M_1\simeq M_2$ [in contrast
to $M_1\simeq 0.5 M_2$ predicted by \Eq{gauginomassrelation}] has
recently been advocated \cite{kane} in order to provide a possible
explanation for the famous CDF $ee\gamma\gamma$ event.
\item Scenarios with $M_2<M_1$, resulting in the $\cnone$ and
$\cpone$ both being wino-like and nearly degenerate in mass.
\item The light gluino scenario, in which the $M_i$ are set to zero.
\item Scenarios in which non-universality for the $M_i$
is such that the gluino has substantial mass, but
is the lightest of the gauginos.
\end{itemize}

\subsubsection{\bf Non-Universal Gaugino Masses}

Non-universal gaugino masses are particularly interesting from
a phenomenological viewpoint.
Models with substantial theoretical motivation include the following.
\bit
\item The O-II Orbifold String model with all matter fields in the $n=-1$ 
untwisted sector, and supersymmetry breaking
dominated by the overall size modulus (as opposed to the dilaton).
The phenomenology of this model is explored in Ref.~\cite{guniondrees2}.
\eit
This is the only known string-based model in which
the limit of pure modulus (no dilaton) supersymmetry breaking is
possible without charge and/or color breaking.
The gaugino masses arise from 1-loop terms generic to any theory;
the resulting $\mgut$ boundary conditions take the form
\begin{equation}
M_a^0\sim \sqrt 3 \mth[-(b_a+\delgs)K\eta],~~m_0^2=\mth^2[-\delgs\kpr],~~
A_0=0\,,
\label{bcs}
\end{equation}
where the $b_a$ are the gauge RGE coefficients
($b_3=3$, $b_2=-1$, $b_1=-33/5$), $\delgs$ is the
Green-Schwarz mixing parameter (a negative integer in the O-II model, with
$\delgs=-3,-4,-5$ preferred), and $\eta=\pm1$.
The one-loop estimates of $K=4.6\times 10^{-4}$ and $\kpr=10^{-3}$ imply
$\mslep,\msq\gg M_a$.
We emphasize that the above relation of $M_a$ to $b_a,\delgs$ 
that arises at one-loop is more general
than the O-II model; in particular, it is likely to survive
non-perturbative corrections. However, the relation between $K$
and $\kpr$ is model-dependent.

\begin{table}[h]
\begin{center}
\begin{normalsize}
\begin{tabular}{|c|ccc|ccc|}
\hline
\ & \multicolumn{3}{c|} {$\mgut$} & \multicolumn{3}{c|}{$\mz$} \cr
$F_{\Phi}$ 
& $M_3$ & $M_2$ & $M_1$ 
& $M_3$ & $M_2$ & $M_1$ \cr
\hline 
${\bf 1}$   & $1$ &$\;\; 1$  &$\;\;1$   & $\sim \;6$ & $\sim \;\;2$ & 
$\sim \;\;1$ \cr
${\bf 24}$  & $2$ &$-3$      & $-1$  & $\sim 12$ & $\sim -6$ & 
$\sim -1$ \cr
${\bf 75}$  & $1$ & $\;\;3$  &$-5$      & $\sim \;6$ & $\sim \;\;6$ & 
$\sim -5$ \cr
${\bf 200}$ & $1$ & $\;\; 2$ & $\;10$   & $\sim \;6$ & $\sim \;\;4$ & 
$\sim \;10$ \cr
\hline
 $\stackrel{\textstyle O-II}{\delgs=-4}$ & $1$ & $\;\;5$ & ${53\over 5}$ & 
$\sim 6$ & $\sim 10$ & $\sim {53\over5}$ \cr
\hline
\end{tabular}
\end{normalsize}
\caption{\baselineskip=0pt $M_a$ at $\mgut$ and $\mz$
for the four $F_{\Phi}$ irreducible representations
and in the O-II model with $\delgs\sim -4$.}
\label{masses}
\end{center}
\end{table}

\begin{table}[h]
\begin{center}
\begin{small}
\begin{tabular}{|c|c|c|c|c|c|}
\hline
\ & ${\bf 1}$ & ${\bf 24}$ & ${\bf 75}$ & ${\bf 200}$ & 
$\stackrel{\textstyle O-II}{\delgs=-4.7}$ \cr
\hline 
$\mgl$          & 285 & 285 & 287 & 288 & 313 \cr
$\msur$         & 302 & 301 & 326 & 394 &  big   \cr
$\mstopone$     & 255 & 257 & 235 & 292 &   big   \cr
$\mstoptwo$     & 315 & 321 & 351 & 325 &   big   \cr
$\msbl$         & 266 & 276 & 307 & 264 &   big   \cr
$\msbr$         & 303 & 303 & 309 & 328 &   big   \cr
$\mslepr$       & 207 & 204 & 280 & 437 &   big   \cr
$\mslepl$       & 216  & 229 & 305 & 313 &   big   \cr
$\mcnone$       & 44.5 & 12.2 & 189 & 174.17  &   303.09 \cr
$\mcntwo$       & 97.0 & 93.6 & 235 & 298 &   337 \cr
$\mcpmone$      & 96.4 & 90.0 & 240 & 174.57 &   303.33 \cr
$\mcpmtwo$      & 275 & 283  & 291 & 311 &    big \cr
$\mhl$          & 67  & 67   & 68   & 70   &   82 \cr
\hline
\end{tabular}
\end{small}
\caption{Sparticle masses for the Snowmass96 comparison point.}
\label{susymasses}
\end{center}
\end{table}


\begin{figure}[h]
\centering
\mbox{%
\epsfig{file=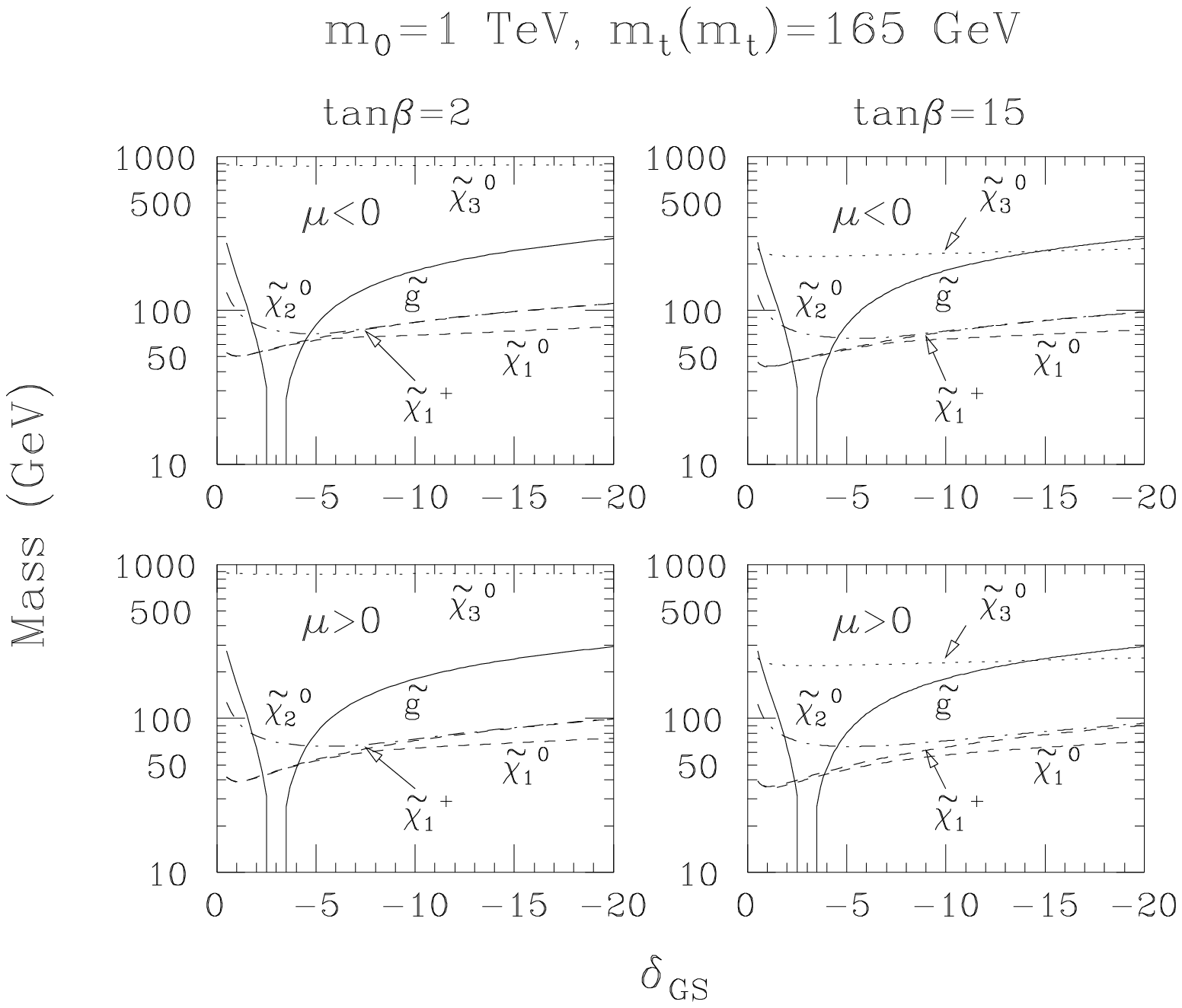,width=3.5in}
}%
\caption{\baselineskip=0pt O-II
model predictions for $\mcnone$, $\mcntwo$, $\mcnthree$, $\mcpone$,
and $\mgl$ as functions of $\delgs$ at $\tanb=2,15$
for $\mu>0$ and $\mu<0$.}
\label{inos}
\end{figure}

\bit
\item $F$-term breaking with $F\neq$ SU(5) singlet. The phenomenology
of these models is discussed in Ref.~\cite{snowtheory2}.
\eit
In these models, it is supposed that the hidden sector
is not completely hidden; it contains some fields with ordinary quantum
numbers. In particular, referring to \Eq{callform},
suppose that $\wh \Phi$ is not a singlet field. In general,
it is only required that
\begin{equation}
\wh \Phi,F_{\Phi}\in ({\bf 24}{\bf \times} 
{\bf 24})_{\rm symmetric}={\bf 1}\oplus {\bf 24} \oplus {\bf 75}
 \oplus {\bf 200}\,.
\label{irrreps}
\end{equation}
For higher representations, only $F_{\Phi}$ components
that are `neutral' with respect to the 
SU(3), SU(2), U(1) gauge groups can acquire a vev if these groups are to remain
unbroken after supersymmetry breaking. The result is that
$\langle F_{\Phi} \rangle_{ab}=c_a\delta_{ab}$, with $c_a$ depending
on the representation (an arbitrary superposition of representations
also being possible). The relative gaugino masses are tabulated in
Table~\ref{masses}.

To obtain a first appreciation for the phenomenological implications
we give in Table~\ref{susymasses} the low-energy gaugino
masses for these cases, assuming the Snowmass96 point $\mgl\sim 285\gev$
value for the gluino mass.
The fact that 
$\mcnone\sim {\rm min}(M_1,M_2)$ and $\mcpmone\sim M_2$ (after mass matrix
diagonalization) implies that in the ${\bf 200}$  and O-II models (which
have $M_2\lsim M_1$)
$\mcpmone\simeq\mcnone$ (and both are winos).
Of course, the relative masses depend upon $\tanb$ as illustrated in
Ref.~\cite{snowtheory2}.
In the O-II model, the dependence of the ino masses on $\delgs$, illustrated in
Fig.~\ref{inos}, is also quite illuminating.  Note that for a narrow
range of $\delgs$ (near the theoretically preferred values), the gluino
will be the LSP! The detailed phenomenological implications of 
these results will be outlined later.

\bit
\item The light-gluino scenario. 
\eit
In this scenario \cite{glennys}, the
soft-supersymmetry-breaking gaugino masses $M_i$ are taken to be
zero at tree-level and only arise at one-loop.  
The result is that $\cnone\sim\wt\gam$ with $\mcnone\sim 0.1-1.5\gev$ (from
one-loop squark-quark and wino/higgsino-Higgs/vector boson diagrams)
and $\mgl \sim 10-600\mev$ (from one-loop squark-quark diagrams). The
lightest supersymmetric particle will be the 
lightest `R-hadron', 
which is likely to be the $R^0=\gl g$ with mass of order 1.5 GeV.
It is claimed \cite{nogluinos}
that this model is highly disfavored on the basis of $2j$ and $4j$
event shapes in $Z$ decay, which, for example,
constrain the running of $\alpha_s$
via $b=(11/6)C_A/C_F-(2/3)n_fT_F/C_F$.  A light $\gl$ acts like 3 new
flavors, increasing $n_f$ from $\sim 5$ to $\sim 8$. Another
immutable prediction of the model is that $\mcpmone\leq \mw$. Thus,
LEP-2 running at $\rts\sim 190\gev$ should soon find evidence for
the chargino (despite the possibly complicated decay patterns) or
the model will be ruled out. It is important to note that
these searches do not
constrain the possibility noted in the previous item of a heavy gluino-LSP.

\subsection{Gauge-mediated supersymmetry-breaking}

The theory of
gauge-mediated supersymmetry breaking (GMSB) \cite{rattazzi} posits that
supersymmetry breaking is transmitted to the MSSM via gauge forces.
Two model-building approaches have been recently explored in the
literature.  First, in hidden-sector models, the GMSB model consists of
three distinct sectors: a hidden (DSB) sector
where supersymmetry is broken, a ``messenger'' sector consisting of
messenger fields with SU(3)$\times$SU(2)$\times$U(1) quantum
numbers, and a sector containing
the fields of the MSSM \cite{dine}.  The
direct coupling of the messengers to the hidden sector generates a
supersymmetry-breaking spectrum in the messenger sector.
Second, in models of ``direct gauge mediation'' (sometimes called
direct-transmission models) \cite{dgm}, the GMSB model consists only of
two distinct sectors: the DSB sector (which also contains the messenger
fields) and the sector of MSSM fields.  In both classes of models,
supersymmetry-breaking is transmitted to the MSSM via
SU(3)$\times$SU(2)$\times$U(1)
gauge interactions between messenger fields
and the MSSM fields.  In particular, supersymmetry-breaking masses for
the gauginos and squared-masses for the squarks and sleptons arise,
respectively, from one-loop and
two-loop diagrams involving the virtual exchange of messenger fields.

In simple hidden-sector
GMSB models, gaugino masses have the same relative values
as if they were unified as specified in \Eq{gunif} (with
$\mx$ of order $\mgut$) even though
the actual initial conditions are set at scale $M_m$.
For example, in the minimal
model where supersymmetry-breaking effects in the messenger sector are
small compared to $M_m$, one finds the following result
for the gaugino Majorana mass terms \cite{dine}:
\beq 
M_i(M_m)=k_i\nmess g\left({\Lambda\over M_m}\right) 
{\alpha_i(M_m)\over 4\pi}\Lambda\,,
\label{gmsbmi}
\eeq
where $k_2=k_3=1,k_1=5/3$, and 
$\nmess$ denotes the effective number of messengers in
the messenger sector --- 
$\nmess\leq 4$ is required to avoid Landau poles.
$\Lambda$ is an effective mass scale that depends on the details
of the GMSB model.  
In addition,
scalar masses are expected to be flavor-independent since the mediating
gauge forces are flavor-blind.  In the simplest 
hidden-sector models they take the form:
\beq
m_i^2(M_m)=2\Lambda^2\nmess f\left({\Lambda\over M_m}\right)
\sum_{i=1}^3 c_i \left({\alpha_i(M_m)\over 4\pi}\right)^2
\label{gmsbmsq}
\eeq
with $c_3=4/3$ (triplets) $c_2=3/4$ (weak doublets), and 
$c_1={5\over 3}\left({Y\over 2}\right)^2$ in 
the normalization where $Y/2=Q-T_3$. 
Thus, degeneracy among families is broken only by effects of order quark or
lepton Yukawa couplings.
In Eqs. (\ref{gmsbmi}) and (\ref{gmsbmsq}), $M_m/\Lambda> 1$ is required
to avoid negative mass-squared for bosonic members of the messenger sector; 
$M_m/\Lambda\geq 1.1$ is
preferred to avoid fine-tuning, for which $f(\Lambda/M_m)\simeq 1$ and
$1\leq g(\Lambda/M_m)\leq 1.23$. For $M_m/\Lambda\geq 2$, 
$1\leq g(\Lambda/M_m)\leq 1.045$. 
In the approximation $g=f=1$, one finds (at tree-level)
$\msq:\mslepl:\mslepr:M_1=11.6:2.5:1.1:\sqrt{\nmess}$,~\footnote{Evolution
corrections \cite{martin} in going from scale $M_m$ to the $\lsim 1\tev$
scale do not destroy this ordering unless $M_m$ is very large, although
they do affect the precise numerical ratios.}
which implies that the next to lightest supersymmetric particle (the
goldstino being the lightest) is
the $\cnone\simeq\wtil B$ for $\nmess=1$ or the $\slepr$'s for $\nmess\geq2$.
Note that supersymmetric scalar and gaugino states with the same color quantum
numbers have masses of roughly the same magnitude.  

\begin{figure}[htb]
\centering
\mbox{%
\epsfig{file=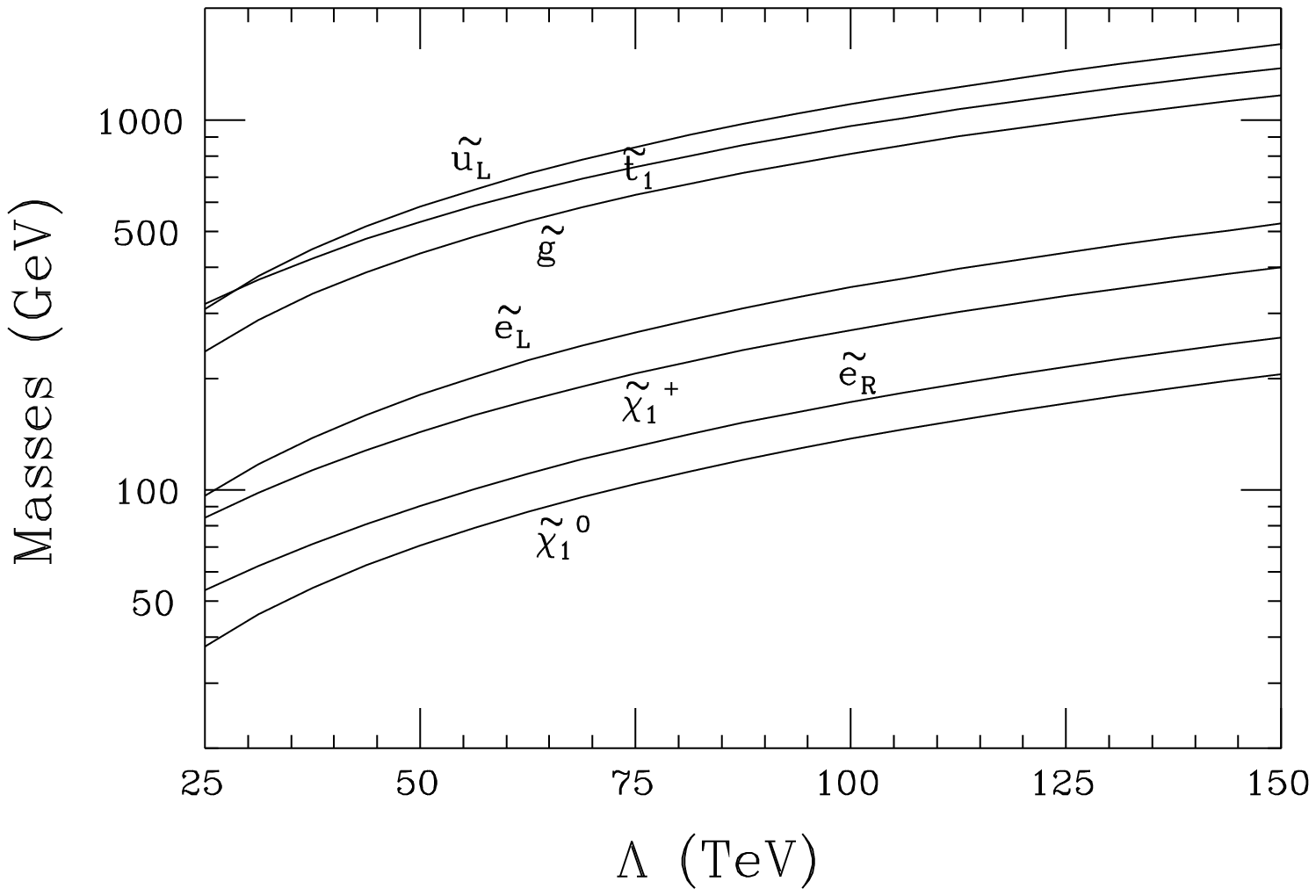,width=3.5in}
}%
\caption{\baselineskip=0pt 
$\mcnone$, $\mcpone$, $\mselr$, $\msell$,
$\mgl$, $\mstopone$ and $m_{\wt u_L}$
as functions of $\Lambda$ in the $\nmess=1$ GMSB scenario,
taking $f=g=1$ and $M_m=1.1\Lambda$.}
\label{gmsbmassfig}
\end{figure}

In order that all
superpartner masses be $\lsim 1\tev$, it follows [\eg, from \Eq{gmsbmi}]
that $\Lambda\lsim 100\tev$. For example, for $\nmess=1$
\beq
\Lambda\sim 80\tev\left({M_1\over 100\gev}\right)\,,
\label{lammassrel}
\eeq
where $M_1$ is the low-energy value of the 
U(1)-gaugino soft-supersymmetry-breaking mass and the gluino mass
is $\mgl\sim 7 M_1$. The masses of various sparticles in the case of $\nmess=1$
are plotted as a function of $\Lambda$ (taking $f=g=1$) in
Fig.~\ref{gmsbmassfig}.

As for the other parameters of low-energy supersymmetry, the
$A$-parameters are suppressed, while the generation of $\mu$ and $b$
is quite model-dependent (and lies somewhat outside the standard
ansatz of
gauge-mediated supersymmetry breaking).  The initial conditions for the
soft-supersymmetry-breaking running parameters are fixed at
$M_m$, which characterizes the average mass of
messenger particles.  In principle, $M_m$ can lie anywhere between
(roughly) $\Lambda$ and $10^{16}$~GeV (in models with larger values of
$M_m$, supergravity-mediated effects would dominate the gauge-mediated
effects). Thus, the minimal GMSB model \cite{thomas} contains 18
Standard Model parameters,
$\Lambda$, $\tanb$ and ${\rm arg}(\mu)$
[after trading in $b$ and $|\mu|^2$ for $v$ and $\tanb$].
There is also a weak logarithmic
dependence on $M_m$, which enters through renormalization group running.
We thus end up with 22 free parameters. Clearly, 
the minimal GMSB approach is even more predictive than the mSUGRA model.

When the GMSB approach is extended to
incorporate gravitational phenomena, supergravity effects will
also contribute to supersymmetry breaking. However, in
models of gauge-mediated supersymmetry breaking, one usually chooses the
model parameters in such a way
that the virtual messenger exchange diagrams dominate the effects of the
direct gravitational interactions between the hidden and visible
sectors. However, of critical importance is the fact
that the super-Higgs effect becomes operative, 
and the gravitino absorbs the goldstino and becomes massive.   
Using \Eq{mgravitino}, it is convenient to rewrite the formula
for the gravitino mass as follows:
\begin{equation}
m_{3/2}={F\over \sqrt 3\mpl}\sim 2.5
\left({\sqrt{F}\over 100\tev}\right)^2~{\rm eV}\,.
\label{mgform}
\end{equation}
The scale of supersymmetry breaking, 
$\sqrt F$, is a very crucial parameter in GMSB models.
It is highly model-dependent.  In hidden-sector models, values
of $\sqrt{F}\gsim 10^3$--$10^4$~TeV are required in order to generate
sufficiently large supersymmetry-breaking in the sector of MSSM
fields \cite{gunionbigf,murayamabigf,chenguniongampub}.
In particular, one can derive \cite{gunionbigf,chenguniongampub} 
the approximate inequality
\beq
\left({\sqrt F\over 2000\tev}\right)\geq f \left({\mslepr\over 100\gev}\right)
\left({1\over 27.5\sqrt{\nmess}}\right)\,,
\label{bigfeq}
\eeq
where $f\sim \left({g_m^2\over 16\pi^2}\right)^{-2}\sim
2.5\times 10^4/g_m^4$, with
$g_m$ being the coupling of the gauge group
responsible for communicating (via two-loop diagrams)
supersymmetry breaking from the
hidden sector to the messenger sector. Perturbativity would require
$g_m\lsim 1$.
For $\mslepr\geq 45\gev$ (the rough LEP limit)
and $f=2.5\times10^4$, $\sqrt F\geq 5200\tev$ ($\sqrt F\geq 2600\tev$)
for $\nmess=1$ ($\nmess=4$).~\footnote{One should allow for a factor
of two or three uncertainty in the approximate lower bound.}
In direct-transmission models, the two-loop communication
between the hidden and messenger sectors is eliminated, $f$ 
in \Eq{bigfeq}\ is effectively of order unity, 
and $\sqrt{F}$ can be as low as 100~TeV in phenomenologically viable models.

Eq.~(\ref{mgform}) implies that the gravitino can be quite light
for moderate $\sqrt F$ values $\geq 100\tev$. For $\sqrt F$
values such that $\mgtino$ is
smaller than $\mcnone,\mslepr$, the $\gtino$ will be the LSP.
An upper bound on $\sqrt F$ then arises
in R-parity conserving GMSB models from the fact that
the gravitino LSP is stable, and thus is also a candidate for dark
matter.  While very light gravitinos (eV masses)
will not contribute significantly to the total mass density of the
universe, if the gravitino LSP is too heavy ($\mgtino\gsim
{\rm few~keV}$) its relic mass density in most early universe
scenarios will be greater than the critical density
\cite{coslimit,murayamabigf} in conflict 
with observation.~\footnote{Although the LSP is an
ideal candidate for being a major component of the dark matter,
one need not use this as a constraint to restrict $\sqrt F$.
It may turn out that the main component of the dark matter
has another source.  Some examples are:
the QCD axion, its supersymmetric partner (the axino)
or the lightest stable particle in the GMSB messenger sector.}

From \Eq{mgform}, this means that $\sqrt F$
values above $\sim 3000\tev$ are disfavored. 
As previously noted, the helicity $\pm\half$ components of $\gtino$
behave approximately like the goldstino.  In contrast to SUGRA
models, $F^{-1}$ is now significantly larger, so that the goldstino
coupling to the particles of the MSSM [\Eq{gtrel}] is significantly
enhanced relative to its SUGRA-model coupling strength.  Thus, in GMSB
models, the gravitino is very likely to be the LSP
and will play a phenomenologically crucial role.

However, minimal GMSB is not a fully realized
model.  The sector of supersymmetry-breaking dynamics can be very
complex, and it is fair to say that no complete model of gauge-mediated
supersymmetry yet exists that is both simple and compelling.
In particular, the various direct transmission models 
are very different from one another and from the simple hidden-sector model.
Relations between sparticle and gravitino masses vary tremendously.
As an extreme example, in the model of Ref.~\cite{raby} the
gluino is the LSP and the gravitino is sufficiently heavy as to be
phenomenologically irrelevant.

\section{Constraints on the gluino-LSP possibility}

We have seen that 
it is possible in both SUGRA  and GMSB models that the 
lightest R-hadron containing the (massive) gluino
is the LSP. Thus, this rather unusual possibility deserves serious
consideration. 

First, there are many limits 
on heavy stable charged particles based on mass spectrometer
searches for heavy isotopes of hydrogen and oxygen and accelerator
fixed target searches for heavy new charged particles. Thus, we assume
that it is the neutral $R_0=\gl g$ that is the LSP. We assume that 
other neutral states such as the $\wtil \rho^0=\gl (u\anti u +d\anti d)$
and charged states such as the $\wtil\rho^+=\gl u\anti d$
(and baryonic equivalents) have short lifetimes for decay to the $R^0$.

Limits on a strongly interacting neutral LSP have been considered in
Refs.~\cite{starkman,mohnus,raby}. There are many differences
in these three treatments.
The assessment of constraints is a very messy subject
and is in a state of flux. Nonetheless, the topic is sufficiently
interesting to warrant a brief overview.

As a first ingredient, we need an estimate for the relic density of $R^0$'s.
Ref.~\cite{mohnus}, using standard freeze-out arguments
and $\sigma_{\rm ann}\sim\alpha_s^2/\mrzero^2$, finds
${n_{R^0}\over n_B}\sim 10^{-3}\mrzero(\tev),$ whereas Ref.~\cite{raby}
obtains ${n_{R^0}\over n_B}\sim 10^{-7}\mrzero(\tev)$. A factor of
$10^3$ in this difference appears to be due to
the much larger annihilation cross section [$\sigma_{\rm ann}\sim
40~{\rm mb}\times \left(m_{\rm proton}/\mrzero\right)^2$] implicitly
assumed in Ref.~\cite{raby}.
Even for the larger relic density of Ref.~\cite{mohnus}, one
finds  ${\rho_{R^0}\over \rho_B}\equiv{n_{R^0}\mrzero\over n_Bm_B}\sim
\left({\mrzero\over1\tev}\right)^2$, implying that
the $R^0$ cannot be the source of dark matter if $\mrzero\lsim 1\tev$.
This result, in turn,  implies that if $\mrzero\lsim 1\tev$ then
it is unlikely that the galactic halo is entirely made of ${R^0}$'s. 

Given a density of $R^0$'s, an important ingredient in obtaining limits is
the flux in the vicinity of the earth of the $R^0$'s.
If the halo density ($\rho=0.4\gev/{\rm cm}^3$)
is dominated by ${R^0}$'s, implying $n_{R^0}=\rho/ \mrzero$,
and using
the velocity of movement of the earth through the halo ($v_{R^0}\sim 2-3 \times
10^7$~cm/s), the flux of ${R^0}$'s from the galactic halo would be
$\phi_{R^0}=n_{R^0}v_{R^0}\sim 10^4 \left({1\tev\over \mrzero}\right) 
{\rm cm}^{-2}{\rm s}^{-1}$. On the other hand, if we employ
the relic density ${n_{R^0}\over n_B}\sim 10^{-3}\mrzero(\tev)$ from
Ref.~\cite{mohnus} with $n_B\sim 2\times 10^{-7}~{\rm cm}^{-3}$,
one finds (using the same $v_{R^0}$)
$\phi_{R^0}\sim 6\times 10^{-3} \left({\mrzero\over 1\tev}\right) 
{\rm cm}^{-2}{\rm s}^{-1}$, \ie\ much smaller for $\mrzero\lsim 1\tev$.

Other important ingredients in obtaining limits on the $R^0$ are 
the $R^0$ cross sections. Ref.~\cite{mohnus} estimates
$\sigma(R^0 p)\sim 10$~mb and $\sigma^{\rm ann}(R^0R^0)\sim
\alpha_s^2/\mrzero^2$ (a larger value applying
if the kinetic energies of the $R^0$'s
are below a GeV or so). Ref.~\cite{raby} suggests that the
leading contribution to
$\sigma(R^0 p)$ would come from glueball exchange: $\sigma(R^0p)\sim
40 (m_\pi/m_{\rm glueball})^2$~mb. Since
the lightest glueball is $\sim 10$ times
heavier than a pion, $\sigma(R^0p)$ would then be $\lsim 0.4$~mb. 

Limits on the $R^0$
derive from: (1) dark-matter (wimp) balloon and underground mine
detection experiments;
(2) capture and co-annihilation in the earth \cite{starkman}
(to produce $\nu$'s detected in proton decay detectors, 
at least as secondary products of primary annihilation)
or capture and co-annihilation in the halo \cite{mohnus} 
to produce energetic $\gam$'s that are detected;
(3) limits on heavy isotopes created by the flux of $R^0$'s 
impacting the earth. We consider each in turn.

(1) Ref.~\cite{mohnus} considers only underground wimp detectors.
First, they argue that since the $R^0$-nuclear cross section is $\gsim 10^{10}$
times larger than the generic heavy neutrino cross section, the wimp
detectors would certainly have detected any $R^0$ flux, even if only
at their predicted `relic' flux level, {\it
if the flux makes it to the detector.}
Then, the critical ingredient is the penetration depth $l_p$ of the $R^0$;
$l_p$ tends to be large because of a small energy transfer per collision. 
Using a geometric estimate of $\sigma(R^0 A)\sim \pi A^{2/3}~{\rm (Fermi)}^3$,
Ref.~\cite{mohnus} estimates $l_p\sim 50~{\rm m}\times \left(
\mrzero/1\tev\right)$.   The result is
that for $\mrzero\lsim 100\gev$, $l_p$ is too small
for the flux to penetrate to any wimp detectors. Of course,
$l_p$ would be further increased if the smaller $\sigma(R^0A)$ 
estimate of \cite{raby} were correct. On the other hand, the applicability
of these limits is more marginal if the relic flux is as small as 
that obtained in Ref.~\cite{raby}. Further, it is important to note
that the effects of energy degradation are estimated to be small 
in Ref.~\cite{mohnus}, whereas they are found to be sufficiently
substantial in Ref.~\cite{starkman} as to reduce rates below observable levels
in all the cases considered there (assuming the fluxes are rescaled
to the relic density level).

(2) The capture and co-annihilation event 
rates depend upon the annihilation cross section and the fraction
of the annihilation product energy appearing in the relevant channel,
as well as whether or not the relevant density is halo or relic.
For both types of observation, the conclusion seems to be that
events would be likely to have been seen if the density is halo-like,
but probably not if relic-like.

(3) Since the $R^0$ can almost certainly 
be captured by oxygen~\footnote{We presume that the
capture rate by protons is much smaller, if not zero.
Otherwise, the much stronger limits on heavy hydrogen isotopes almost
certainly rule out the $R^0$ LSP.}
limits on exotic oxygen isotopes contained in water molecules
in the ocean will place limits on the $R^0$. 
The limits on such heavy water
given in Ref.~\cite{mohnus} require
$\phi_{R^0}\lsim 0.1 \mrzero(\tev) \left({{\rm 100~Myr}
\over t_{\rm acc}}\right){\rm cm}^{-2}{\rm s}^{-1}$, 
where $t_{\rm acc}$ is the
accumulation time, with 100 Myr being an estimate for the
age of the ocean. This flux limit can be compared with the fluxes
computed earlier. One observes that
if $R^0$'s are the dominant halo component, then we would require
$\mrzero>100\tev$. However, we have already argued that
the halo density should not be dominated by $R^0$'s. The $R^0$
flux estimated using the Ref.~\cite{mohnus} relic $R^0$ density prediction
is well below the above limit.
In Ref.~\cite{raby} the estimated relic $R^0$ density is even smaller,
and the dominant source of $R^0$'s impinging
on the ocean is $R^0$ production in the atmosphere by cosmic rays.
The crude estimate of the heavy isotope abundance arising from
this source borders on the observed limit. 

In summary, it appears that we can escape the above limits 
for $\mrzero\lsim 1\tev$ so long as the halo density
of $R^0$'s is similar to the (very small) relic density 
estimated in Ref.~\cite{raby}. Alternatively, a relic
density as large as that estimated in Ref.~\cite{mohnus} is
allowed if energy degradation effects are much larger than
allowed for in their wimp detection estimates or if $\mrzero\lsim 100\gev$.
In either case, the $R^0$'s cannot be a significant source of dark matter.

\section{Beyond the MSSM}

Non-minimal models of low-energy supersymmetry can also be constructed.
One approach is to add new structure beyond the Standard
Model at the TeV scale or below.  The supersymmetric extension of such a
theory would be a non-minimal extension of the MSSM.  Possible new structures
include: (i) the supersymmetric generalization of the see-saw model of
neutrino masses;
(ii) an enlarged electroweak gauge group beyond SU(2)$\times$U(1);
(iii) the addition of new, possibly exotic, matter
multiplets [{\it e.g.}, a vector-like color triplet with electric charge
$\third e$; such states sometimes occur as low-energy remnants in E$_6$
grand unification models]; (iv) the addition of low-energy
SU(3)$\times$SU(2)$\times$U(1) singlets; and/or (v) the addition
of one or more new families with massive neutrinos.

A second approach is to retain the minimal particle content of the MSSM
but remove the assumption of R-parity invariance \cite{dreiner}.
The most general explicit~\footnote{The phenomenology
of models in which there is spontaneous R-parity violation from
a sneutrino field vacuum expectation value is described elsewhere in these
proceedings.}
R-parity-violating (RPV) theory involving the MSSM spectrum introduces
many new parameters to both the supersymmetry-conserving and the
supersymmetry-breaking sectors.  Each new interaction term
violates either B or L conservation.   For example, new
scalar-fermion Yukawa couplings would be:
\beq
(\lambda_L)_{pmn} \widehat L_p \widehat L_m \widehat E^c_n
+ (\lambda_L^\prime)_{pmn}\widehat L_p \widehat Q_m\widehat D^c_n
+(\lambda_B)_{pmn}\widehat U^c_p \widehat D^c_m \widehat D^c_n\,,
\label{rpv}
\eeq
where $p$, $m$, and $n$ are generation indices, and
gauge group indices are suppressed.  In the notation
above, $\wh Q$, $\wh U^c$, $\wh D^c$, $\wh L$, and $\wh E^c$
respectively represent
$(u, d)_L$, $u^c_L$, $d^c_L$, $(\nu$, $e^-)_L$, and $e^c_L$ and the
corresponding superpartners.  The Yukawa interactions are obtained
from \Eq{rpv}\ by
taking all possible combinations involving two fermions
and one scalar superpartner.  Note that the term in \Eq{rpv}\
proportional to $\lambda_B$ violates B, while the other two terms
violate L.

Additional R-parity violating effects arise from the mixing of
$\hat L_p$ and the $Y=-1$ Higgs superfield $\hat H_d$.  This leads to
new parameters $\mu^\prime_p$ ($p=1,2,3$) which are analogues of
the supersymmetric Higgs mass parameter $\mu$.  Finally, there are
soft-supersymmetry-breaking R-parity-violating $A$-terms and a $B$-term
contributing to the scalar potential (\ie, the interaction of squarks,
sleptons and Higgs bosons):
\beq
(a_L)_{pmn} \widetilde L_p \widetilde L_m \widetilde E^c_n
+ (a_L^\prime)_{pmn}\widetilde L_p \widetilde Q_m\widetilde D^c_n
+(a_B)_{pmn}\widetilde U^c_p \widetilde D^c_m \widetilde D^c_n
+b'_p \widetilde L_p H_u\,,
\label{rpvaterms}
\eeq
where $H_u$ is the $Y=1$ doublet of Higgs scalars.

Phenomenological constraints on
various low-energy L and B-violating processes yield limits on each of
the coefficients $(\lambda_L)_{pmn}$, $(\lambda_L^\prime)_{pmn}$ and
$(\lambda_B)_{pmn}$ taken one at a time \cite{dreiner}.
If more than one coefficient is simultaneously non-zero, then the limits
are in general more complicated.  All possible RPV terms
cannot be simultaneously present and unsuppressed; otherwise
the proton decay rate would be many orders of magnitude larger than
the present experimental bound.   One way to avoid proton decay is to
impose B or L-invariance (either one alone would suffice).
For example, models of ``baryon-parity'' possess a discrete symmetry
that requires $\lambda_B=0$ but allows non-zero L-violating terms in
\Eq{rpv}.

\section{The Higgs sector of low-energy supersymmetric models}

In the MSSM, the Higgs sector is a two-Higgs-doublet model
with Higgs self-interactions constrained by supersymmetry
\cite{Gunion90,Gunion97}.
After absorbing unphysical phases
into the definition of the Higgs fields one finds that the Higgs
sector is CP-conserving, so that
$\tan\beta$ is a real parameter (conventionally chosen to be positive).
The physical neutral Higgs scalars are CP-eigenstates.
The five physical Higgs particles are those listed earlier.
At tree level, $\tan\beta$
and one Higgs mass (usually chosen to be $\mha$)
determine the tree-level Higgs-sector parameters.
These include the other Higgs masses,
an angle $\alpha$ [which measures the component of the original
$Y=\pm 1$ Higgs doublet states in the physical CP-even
neutral scalars], the Higgs boson self-couplings,
and the Higgs boson couplings to particles of the Standard Model and
their superpartners.

When one-loop radiative corrections are incorporated, the Higgs
masses and couplings depend on
additional parameters of the supersymmetric model
that enter via virtual loops.  The impact of these corrections
can be significant. Most importantly,
the tree-level MSSM-124 upper bound of 
$\mhl\leq m_Z|\cos 2\beta|\leq m_Z$ is increased to
$\mhl\lsim 125-130\gev$
for $m_t=175$~GeV and a top-squark mass of $\mstop\lsim 1\tev$.
The charged Higgs mass is also constrained in the MSSM.  At tree level,
$\mhpm^2=\mw^2+\mha^2$, which implies that charged Higgs bosons cannot
be pair produced at LEP-2.  Radiative corrections modify the tree-level
prediction, but the corrections are typically small.

In the parameter regime of $\mha\gsim 2\mz$, one finds that
the couplings of the $\hl$ are nearly
indistinguishable from those of the Standard Model Higgs boson.
Moreover, the non-minimal Higgs states, $\hh$, $\ha$, and $\hpm$ are heavy and
approximately degenerate in mass.  This regime, called the {\it decoupling
limit}, is rather generic and applies also to models with more general
Higgs sectors. Present experimental bounds
do not yet require the Higgs sector parameters to lie in the
decoupling regime.  In this regard, one of
the most sensitive observables is the decay rate for $b\to s\gamma$.
Current data yields a rate that is within $2\sigma$ of the Standard
Model prediction.
Applying these results to the MSSM, the observed rate (which is {\it
below} the Standard Model prediction) would require a negative one-loop
contribution of a light chargino and top-squark to cancel the positive loop
contributions of the $W^\pm$ and $\hpm$.
In the absence of light supersymmetric particle loops, one
obtains a strong lower bound on the charged Higgs mass, since the latter
contribution raises the predicted rate for $b\to s\gamma$.  In this
case, one finds that $\mha\gsim 350\gev$, a result well into
the decoupling regime.

The MSSM Higgs mass bound quoted above does not in general apply to
non-minimal supersymmetric extensions of the Standard Model.
If additional Higgs singlet and/or triplet fields are introduced,
then new Higgs self-coupling parameters appear, which are
not significantly constrained by present data.  These parameters can
contribute to the light Higgs masses; the upper bound on these
contributions depends on an extra assumption beyond the physics
of the TeV-scale effective theory.  For example, in the
simplest non-minimal supersymmetric extension of the
Standard Model (NMSSM),
the addition of a Higgs singlet adds a new Higgs
self-coupling parameter, $\lambda$ \cite{singlets}.
The mass of the lightest neutral Higgs boson can be
raised arbitrarily by increasing the value of $\lambda$.
Under the assumption that all couplings stay perturbative
up to the Planck scale, one finds
in almost all cases that $\mhl\lsim 150$~GeV, independent of the
details of the low-energy supersymmetric model
\cite{GKW}.  The NMSSM also permits a tree-level charged Higgs
mass below $\mw$.
However, as in the MSSM, the charged Higgs boson
becomes heavy and approximately degenerate in mass with $\ha$
in the decoupling limit (where $\mha\gg\mz$).

Finally, in
models of R-parity violating supersymmetry, the distinction between the
Higgs and matter multiplets is lost. Thus, R-parity violation permits the
mixing of sleptons and Higgs bosons.  The associated phenomenology will not be
addressed here; see Ref.~\cite{rpvhiggs} for further discussion.

\subsection{Supersymmetric Higgs searches at future colliders}

We now consider the opportunities for detecting the Higgs bosons of R-parity
conserving low-energy supersymmetry at future colliders.
Over (nearly) the complete range MSSM
Higgs sector parameter space, at least one of the MSSM Higgs bosons
will be detectable at the LHC (if not at LEP-2 or the Tevatron), 
assuming that the machine runs at its design
luminosity of $100\pbi$ per year, and under the assumption that the
current detector design capabilities are achieved.
An $\epem$ collider (denoted $e$C) or a $\mu^+\mu^-$ collider
(denoted $\mu$C) would also guarantee detection
of at least one Higgs boson for
all of the MSSM Higgs sector parameter space
(by extending the LEP-2 Higgs search) once the
center-of-mass energy of the machine is above 300~GeV.

\begin{figure}[h]
\centering
\mbox{\epsfig{file=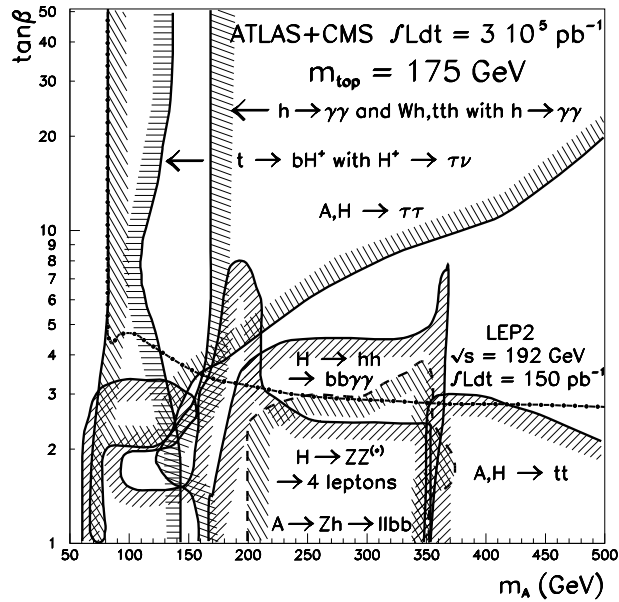,width=3.5in}}
\caption{Discovery contours ($5\sigma$) in the parameter space of the 
minimal supersymmetric model for ATLAS+CMS at the
LHC: $L=300\fbi$. \Twoloop\ radiative corrections 
to the MSSM Higgs sector are included
assuming $\mstop=1\tev$ and no squark mixing.}
\label{mssmhilum2}
\end{figure}

However, complete coverage of the MSSM parameter space only means that
at least one Higgs boson can be detected.  For example, in the
decoupling limit (where $\mha\gsim 2\mz$), $\hl$ will surely be
discovered at LEP-2 or the LHC (and can be
observed at the $e$C).  But, detection of the heavier
non-minimal Higgs states $\hh$, $\ha$, and/or $\hpm$
is not guaranteed.  At the LHC, there is a region \cite{latestplots},
illustrated in Fig.~\ref{mssmhilum2}, in the $(\mha,\tanb)$ parameter space
characterized by $\tanb\gsim 3$ and $\mha\gsim
300\gev$~\footnote{This is probably the only
region of Fig.~\ref{mssmhilum2} that is consistent 
with the measured $\br(b\to s\gam)$ given that $\mstop\sim 1\tev$ is assumed.} 
such that only the $\hl$ will be detectable.~\footnote{
In the regime $1\lsim\tanb\lsim 3$ and $\mha\gsim
300\gev$, Ref.~\cite{latestplots} asserts that $\ha$ and $\hh$ can be
discovered at the LHC via their $t\bar t$ decay mode.  However, there
has not been a full detector simulation and complete analysis of
backgrounds to substantiate such a claim.}
At the $e$C, $\hh\ha$ and $\hp\hm$
pair production is not kinematically allowed if $\mha\gsim \rts/2$.
Moreover, the production rate for $\hl\ha$
(although it may be kinematically allowed) is suppressed in the decoupling
limit.  Thus, the non-minimal Higgs states are
not detectable at the $e$C for $\mha\gsim \rts/2$.

Recently, two important questions regarding the observation of
the Higgs sector of a supersymmetric model have been addressed.

\bit
\item 
First, is discovery of one Higgs boson guaranteed if singlet Higgs fields
are added to the MSSM?
\eit

1. It has been demonstrated \cite{kimoh} that
at least one of the light CP-even Higgs bosons ($\h_{1,2,3}$)
of a two-doublet plus
one singlet NMSSM Higgs sector will be observed at the $e$C. This result
follows from the fact that if the lightest ($\h_1$) has weak $ZZ$
coupling then (one of) the heavier ones must be almost as light
and have substantial $ZZ$ coupling (and thus be discoverable
in the $\epem\to\zstar\to Z\h$ production mode). This result
probably extends to models that contain several singlet Higgs fields.

2. However, before the $e$C 
is built, only data from LEP-2, the Tevatron and the LHC
will be available. One finds \cite{ghm} that 
discovery of at least one Higgs boson is no
longer guaranteed if a singlet Higgs field is present in
addition to the minimal two doublet fields. The
no-discovery ``holes'' in parameter space are never terribly large,
but are certainly not insignificant for moderate values of
$\tanb\sim 3$--10.

In order to demonstrate point 2. above, one employs
the same detection modes for the NMSSM as employed for Fig.~\ref{mssmhilum2}
in the MSSM case:
(1) $\zstar\to Z\h$ at LEP-2; (2) $\zstar\to \h\a$ at LEP-2;
(3) $gg\to \h\to\gam\gam$ at the LHC; 
(4) $gg\to\h\to Z\zstar~{\rm or}~ZZ\to 4\ell$
at LHC; (5) $t\to\hp b$ at the LHC;
(6) $gg\to b\anti b \h,b\anti b\a \to b\anti b \tauptaum$ at the LHC;
(7)  $gg\to\h,\a\to\tauptaum$ at the LHC.
Additional Higgs decay modes that could be considered, but not
reliably, at the LHC include:
(a) $\a\to Z\h$; (b) $\h\to\a\a$;
(c) $\h_j\to\h_i\h_i$; (d) $\a,\h\to t\anti t$.  In order to avoid having to
consider these latter, we explicitly exclude
any choice of NMSSM parameters for which (a)-(d) might be relevant.

The free parameters specifying the Higgs sector
of the NMSSM are: $\tanb$, the mass of the lightest CP-even
Higgs mass eigenstate $\mhi$, the mass of the lightest CP-odd
scalar $\ma$ (we consider only the parameter sub-region in which
the 2nd CP-odd scalar is taken to be much heavier),
the new superpotential coupling $\lam$ appearing in $W\ni \lam \wh H_d \wh H_u
\wh N$, and three mixing
angles $\alpha_{1,2,3}$ which parameterize the orthogonal matrix that
diagonalizes the CP-even Higgs mass-squared matrix. One finds that
$\lam\lsim 0.7$ is required if $\lam$ is to remain perturbative
during evolution from $\mz$ to $\mgut$, which implies
a $\tanb$-dependent upper limit on $\mhi$ below $\sim 140\gev$.

\begin{figure}[htb]
\centering
\mbox{\epsfig{file=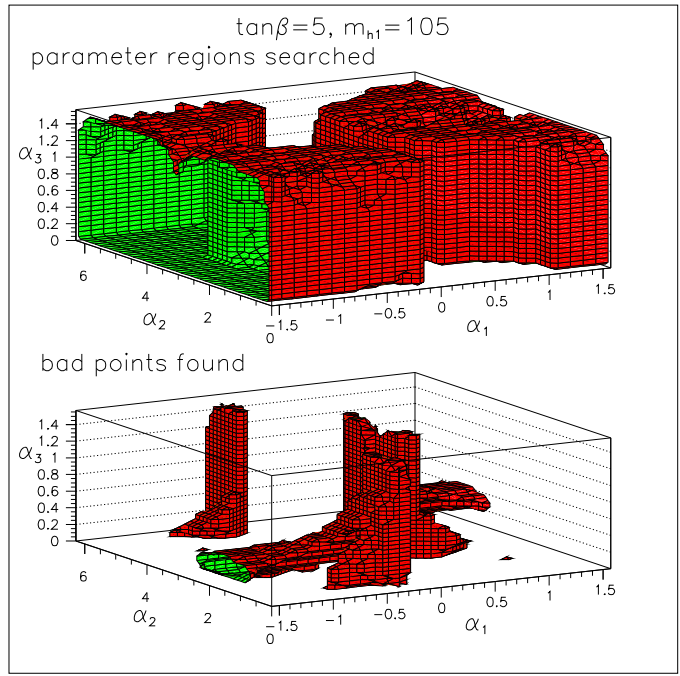,width=3.5in}}
\caption{For $\tanb=5$ and $\mhi=105\gev$, we display in three dimensional
$(\alpha_1,\alpha_2,\alpha_3)$ parameter space the parameter regions
searched (which lie within the surfaces shown), and the
regions therein for which the remaining model parameters can
be chosen so that no Higgs boson is observable
(interior to the surfaces shown).}
\label{tanb5}
\end{figure}

As already noted, moderate $\tanb$ yields the worst case: 
for low $\tanb\leq 1.5$, at least one of the CP-even Higgs is observable
in the standard $\gam\gam,4\ell,\ldots$ modes; at high $\tanb\geq 10-11$, the
$b\anti b \h$ and $b\anti b\a$ modes (with $\h,\a$ decaying to $\tauptaum$)
are sufficiently enhanced to be observable. A typical unobservable
region of $(\alpha_1,\alpha_2,\alpha_3)$ parameter space that arises
at moderate $\tanb$ is illustrated in Fig.~\ref{tanb5}.

3. At a $\mupmum$ collider, direct $s$-channel production $\mupmum\to \h_1$
is quite likely to be visible \cite{bbghunpub}.
Only the (predicted) $\mupmum\h_1$ coupling is needed; the $\h_1$ can
be decoupled from $ZZ,WW$ without affecting its detectability.
If the $\h_1$ has not been previously detected, then
a scan search in the $m_{\h_1}\leq 150\gev$ region is required with
$\Delta E_{\rm beam}/E_{\rm beam}\lsim 0.01\%$. The time required
for this scan could run to many years, depending upon the luminosity
of the muon-collider ($\mu$C) and the size of the $\mupmum\h_1$ coupling.

\bit 
\item A second important question is: 
What is required in order to discover the $\hh,\ha,\hpm$
of the MSSM?  
\eit

Here, a region of particular
concern is the decoupling regime of $\mha\gsim 200\gev$,
where finding evidence for a non-minimal Higgs sector will
be the most problematical. This concern is most pronounced if
$\tanb\gsim 3$, \ie\ such that the LHC cannot observe the $\hh,\ha$
(see above).

1. At the $e$C, the only means for detecting
any of the heavier non-minimal Higgs states
$\hh,\ha,\hpm$ when $\mha\gsim\rts/2$
is to run the collider in
the $\gam\gam$ collider mode \cite{ghgamgam,bbc}.
In this way, single $\gam\gam\to\hh,\ha$
production can be observed for $\mha\lsim 0.8\rts$ provided high
integrated luminosity (\eg, $L\sim 200\fbi$) is accumulated.
There is still no certainty that a $\gam\gam$ collider facility will
be included in the final $e$C plans, nor any guarantee
that the required luminosity can be achieved.

2. At a $\mupmum$ collider, direct $s$-channel production
$\mupmum\to\hh,\ha$ may be observable \cite{bbgh}.
Assuming no constraints on or knowledge of $\mha$,
one would ideally wish to scan
the entire region of $\mha$ between $\rts/2$ (the pair
production limit) and the maximum available $\rts$ and be certain
of seeing the $\hh,\ha$ signal (if present) for any $\tanb\gsim 3$ (the value
below which the LHC will find the $\hh,\ha$).
For $\rts\sim 500\gev$,
this would require an integrated luminosity of $L=100\fbi$ (at modest beam
energy resolution of 0.1\%).
However, current $\mu$C designs suggest that
only $L\lsim 10 \fbi/{\rm yr}$ will be achieved in this energy range.
Thus, if we imagine devoting a period of several years to this scan,
$\hh,\ha$ discovery at the $\mu$C
would only be certain if $\tanb\gsim 5-6$.
As described shortly, the situation would be greatly improved if 
the $\hl$ has been detected and its properties measured at the $e$C
(operating at $\rts=500\gev$) and/or the $\mu$C (operating
in the $s$-channel Higgs production mode). If only $e$C data
is available, it would be possible to distinguish between the $\hl$
and $\hsm$ and at least roughly determine $\mha$ so long as $\mha$
is below $450\gev$. If we only have 
$s$-channel $\mu$C data the result is similar.
If both types of data are available, a rough $\mha$ determination would
be possible up to $\mha\sim 600\gev$. Precision measurements
of the SM-like Higgs properties would thus allow a much narrower scan
for $\hh,\ha$ discovery; as a result,
the $\hh,\ha$ could be detected at the $\mu$C with only one or two
years running for essentially any $\tanb\gsim 1-2$.

3. At both the $\epem$ and $\mupmum$ colliders, it is envisioned that
the machine will be upgraded to increasingly larger $\rts$. Once
$\rts\gsim 2\mha$,
$\hh\ha$ and $\hp\hm$ pair production will be observable.  It has
been shown that this will be true even if these Higgs bosons have
substantial decays to supersymmetric particles \cite{gk,fengmoroi}.

Returning to the MSSM Higgs sector, there are a number of interesting issues.
\bit
\item How precisely will we know $\mhl$ and how useful will it be?
\eit

Certainly $\mhl$ will be very precisely measured:
$\Delta\mhl\lsim 100\mev$ or better is achievable
at the $e$C and LHC; the $\mu$C yields $\Delta\mhl\lsim 1\mev$ 
via an $s$-channel scan. Precise knowledge of
$\mhl$ has the potential to inaugurate a new precision game.
At one-loop (neglecting mixing, \etc.),
$\Delta\mhl^2={3g^2\mt^4\over 8\pi^2\mw^2}\ln {\mstop^2\over\mt^2}$.
For $\mhl=100\gev$, $\Delta\mhl=\pm 100\mev$ implies constraints
of order $\Delta\mt=\pm170\mev$,
$\Delta\mstop=\pm 2\gev$. (More generally and at two-loops, 
$A$ and $\mu$ parameters enter.)
This level of constraint on $\mstop$, $\ldots$ probably exceeds
that achievable experimentally by direct measurements. 
However, there is an important caveat:
can theoretical predictions for $\mhl$ be made
at the $\Delta\mhl=\pm 100\mev$ level? 
The currently available two-loop RGE-improved result 
is only reliable to a few GeV. 

\bit
\item 
What are the prospects for and means for distinguishing the MSSM $\hl$ 
or other SM-like light Higgs bosons from the SM $\hsm$
via branching ratios and couplings, and what are the implications?
\eit

We can only present a very brief summary. Further details
can be found in \cite{snowmasssummary} and \cite{balholm}.
The Snowmass96 study \cite{snowmasssummary} shows that LEP-2, TeV-33
and LHC data will not provide determinations of the Higgs couplings
of sufficient accuracy to distinguish the $\hl$ from the $\hsm$
in the `decoupling' regime of $\mha\gsim 2\mz$.
For example, for a SM-like Higgs mass $\sim \mz$ 
(the best case where data from all
three machines will be available) the fundamental properties that
can be determined (which is a limited subset) and
their errors are:
 $\br(b\anti b)$ --- $\pm 26\%$;
 $(WW\hsm)^2/(ZZ\hsm)^2$ --- $\pm 14\%$;
 $(WW\hsm)^2$ --- $\pm 20\%$;
 $(ZZ\hsm)^2$ --- $\pm 22\%$;
 $(\gam\gam\hsm)^2/(b\anti b\hsm)^2$ --- $\pm 17\%$;
 $\br(\gam\gam)$ --- $\pm 31\%$;
 $(gg\hsm)^2$ --- $\pm 31\%$;
 $(t\anti t\hsm)^2/(WW\hsm)^2$ --- $\pm 21\%$;
 $(t\anti t\hsm)^2$ --- $\pm 30\%$.
At higher masses, LEP-2 data will not be available and TeV-33 
statistics decline rapidly.
Overall, without $e$C and/or $\mu$C data, 
we will know that the Higgs is SM-like, but we will not
be able to discriminate between different SM-like possibilities.

\begin{figure}[p]
\centering
\mbox{\epsfig{file=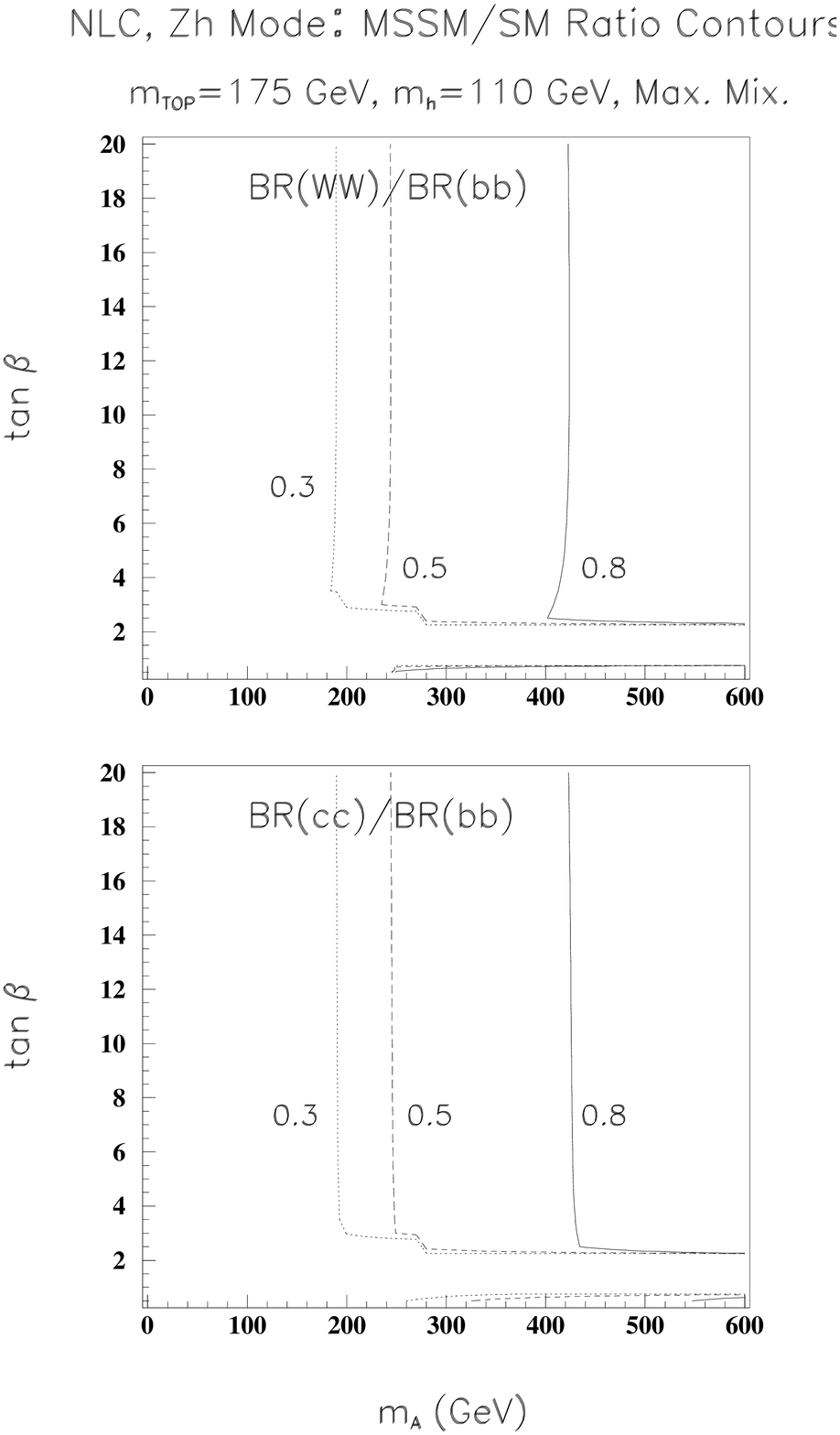,width=3.2in}}
\caption{Constant value contours in $(\mha,\tanb)$ parameter space
for the ratios $[W\wstar/b\anti b]_{\hl}/[W\wstar/b\anti b]_{\hsm}$  and
$[c\anti c/b\anti b]_{\hl}/[c\anti c/b\anti b]_{\hsm}$.
We assume ``maximal mixing'' in the squark sector and present
results for the case of fixed $\mhl=\mhsm=110\gev$. The band extending
out to large $\mha$ at $\tanb\sim 2$ is where $\mhl=110\gev$
is theoretically disallowed in the case of maximal mixing.
For no mixing, the vertical contours
are essentially identical --- only the size of the disallowed band
changes. The same contours apply if $b\anti b$ is replaced by $\tauptaum$.}
\label{figdevsm}
\end{figure}

As a next step, we imagine that we have
data available from the $e$C as well as the LHC.
This yields a dramatic improvement in our ability to determine
the fundamental couplings of the $\hl$.
One finds \cite{snowmasssummary} that
the best `bet' for discriminating between SM-like Higgs
bosons for $\mh\leq 130\gev$ is via the ratio 
$\sigma\br(\h\to c\anti c)/\sigma\br(\h\to b\anti b)$;
for $\mh\geq 130\gev$ (as possibly relevant in the NMSSM), 
$\sigma\br(\h\to W\wstar)/\sigma\br(\h\to b\anti b)$ 
will be the most valuable ratio. 
The results from Snowmass96 use `topological tagging' (\eg\
primary, secondary, tertiary vertices for a
$b$ jet but only primary and secondary vertices for a
$c$ jet).  
For $L=200\fbi$ at $\rts=500\gev$ and $\mh\leq 130\gev$
(and combining $Z\h$, $ZZ$-fusion and $WW$-fusion
production processes), one obtains a statistical error 
in $\br(c\anti c)/\br(b\anti b)$ of $\sim \pm 7\%$.
Theoretical uncertainty in $m_c(m_c)$ and $\mb(\mb)$ (sum rules,
lattice) and QCD running should reach the $< 10\%$ level in a few years,
implying a net error of $\lsim 10\%$. The result is that one
would be able to distinguish $\hl$ from $\hsm$ at 
a $\geq 2\sigma$ level for $\mha\leq
450\gev$ for the typical $\mhl=\mhsm=110\gev$ mass case.
This is illustrated in Fig.~\ref{figdevsm}, where it is seen
that deviations are almost independent of $\tanb$ and 
are larger than $\sim 20\%$ for $\mha\leq 450\gev$.

For the NMSSM, the determination of the $(ZZ\h)^2$ coupling squared could
be very valuable for checking the sum rule
$\sum_i (ZZ\h_i)^2=\sum_i (WW\h_i)^2=1$,
(where the $(VV\h_i)^2$ -- 
$V=W,Z$ -- are defined relative to the SM-values).~\footnote{
$(ZZ\h)^2$ is also crucial to getting many quantities in $e$C only analysis.}
A direct measurement (independent of the $\h$ branching ratios)
is possible using $\epem\to Z\h$ with $Z\to \epem,\mupmum$ and $\epem\to\epem
\h$ ($ZZ$-fusion) production 
and isolation of the $\h$ as a peak in the missing mass
recoiling against the $\ell^+\ell^-$.
Inclusion of the $ZZ$-fusion mode \cite{ghs} as well as the usual $Z\h$
associated production mode, yields $(ZZ\h)^2$
errors in the $\mh<150\gev$ range of order $\pm 5\%-\pm7\%$.
Determining $(WW\h)^2$ is more involved. For example,
for $\mh\leq 140\gev$, one must measure $\sigma(\nu\anti\nu
\h)\br(\h\to b\anti b)$, and divide by $\br(\h\to b\anti b)$.
One can determine $\br(\h\to b\anti b)$, from the
ratios of the type $\sigma(Z\h)\br(\h\to b\anti b)/\sigma(Z\h)$ and
the $ZZ$ fusion analogue, with an accuracy of about 5\%
for $\mh\lsim 130\gev$ (assuming a SM-like $\h$).
The result is
$\sim\pm 5\%-\pm6\%$ error for $(WW\h)^2$ when $\mh\leq 140\gev$.

Determination of $\gamh$ (as needed needed to compute
$(X\h)^2=\br(\h\to X)\gamh$ for $X=b\anti b$, \etc)
requires a lengthy indirect procedure for $\mh\leq 130\gev$.
One needs to determine $\br(\h\to \gam\gam)$ as well as possible and use
$\gam\gam$ collisions to determine $\Gamma(\h\to \gam\gam)$ and then compute
$\gamh={\Gamma(\h\to\gam\gam)\over \br(\h\to\gam\gam)}$.
Unfortunately, $\br(\h\to\gam\gam)$ cannot be well measured at $e$C alone. 
The best results are obtained by
using a complicated procedure involving LHC and $e$C data, as described
in Ref.~\cite{snowmasssummary}. The net result is an
$\br(\h\to\gam\gam)$ error of $\lsim \pm 16\%$ for $\mh\lsim 130\gev$.
To measure $\Gamma(\hsm\to\gam\gam)$, the $\gam\gam$
collider is required. Current estimates,
including systematics, suggest a
$\Gamma(\hsm\to\gam\gam)\br(\hsm\to b\anti b)$ error (for $L=50\fbi$
at the $\gam\gam$ collider)
of 8\%-10\%. Given $\pm5\%-\pm6\%$ error for $\br(\hsm\to b\anti b)$
one finds a
$\Gamma(\hsm\to\gam\gam)$ error of order $\pm 12\%$.
Putting all this together yields
a $\gamhsm$ error of $\sim \pm 18\%$; not very wonderful.
A detailed summary of errors for branching ratios, couplings and
the total width of a SM-like Higgs obtained 
by combining LHC, $e$C and $\gam\gam$
collider data appears in Ref.~\cite{snowmasssummary}.

Production of the $\h$ in the $s$-channel at
a muon collider can provide even greater discrimination power between
the $\hl$ and the $\hsm$, provided $\mh\neq\mz$ \cite{bbgh}.
The crucial measurements are two:
(1) The very tiny Higgs width:
$\gamh=1-10\mev$ for a SM-like Higgs with $\mh\lsim 140\gev$
(i.e. mass as predicted for the $\hl$ of the MSSM).
(2) $\sigma(\mupmum\to \h\to X)$ for 
$X=\tauptaum,c\anti c, b\anti b,W\wstar,Z\zstar$.~\footnote{Note
that $\sigma(\mupmum\to\h\to X)$ provides
a determination of $\Gamma(\h\to\mupmum)\br(\h\to X)$ unless $\sigrts\ll
\gamh$.}
The accuracies estimated to be achievable during an optimized 3-point scan
with total integrated luminosity of $L=0.4\fbi$
(as currently projected for four years of operation
with a beam energy resolution of $R=0.003\%$)
in the case of $\mhl=110\gev$ are the following: 
$\tauptaum$ --- $\pm 8\%$; $c\anti c$ --- $\pm 19\%$;
$b\anti b$ --- $\pm 3\%$;
$W\wstar$ --- $\pm 15\%$; $Z\zstar$ --- $\pm 190\%$; $\gamhsm$ --- $\pm 16\%$.
The important ratios for discriminating $\hl$ from $\hsm$, and their errors, 
are:
${W\wstar\over \tauptaum}\to\pm18\%$,
${c\anti  c\over \tauptaum}\to\pm22\%$,
${W\wstar \over b\anti b}\to\pm15\%$,
${c\anti c\over b\anti b}\to\pm20\%$.
From Fig.~\ref{figdevsm}, these errors imply that with $L=0.4\fbi$
at the $\mu$C one can
distinguish between the $\hsm$ and the $\hl$
at a $>2\sigma$ level for $\mha$ up to $\sim 400\gev$,
which is similar to the discrimination power of the
$e$C operating at $\rts=500\gev$ for $L=200\fbi$.
Note that since the $\gamh$ errors are big and since $\gamh$ is 
model-dependent,  $\gamh$ is not so clearly useful as the
above ratios. Still, deviations of $\gamhl$ from $\gamhsm$ are
predicted to be substantial if $\mha\lsim 500\gev$.

\begin{figure}[h]
\centering 
\mbox{\epsfig{file=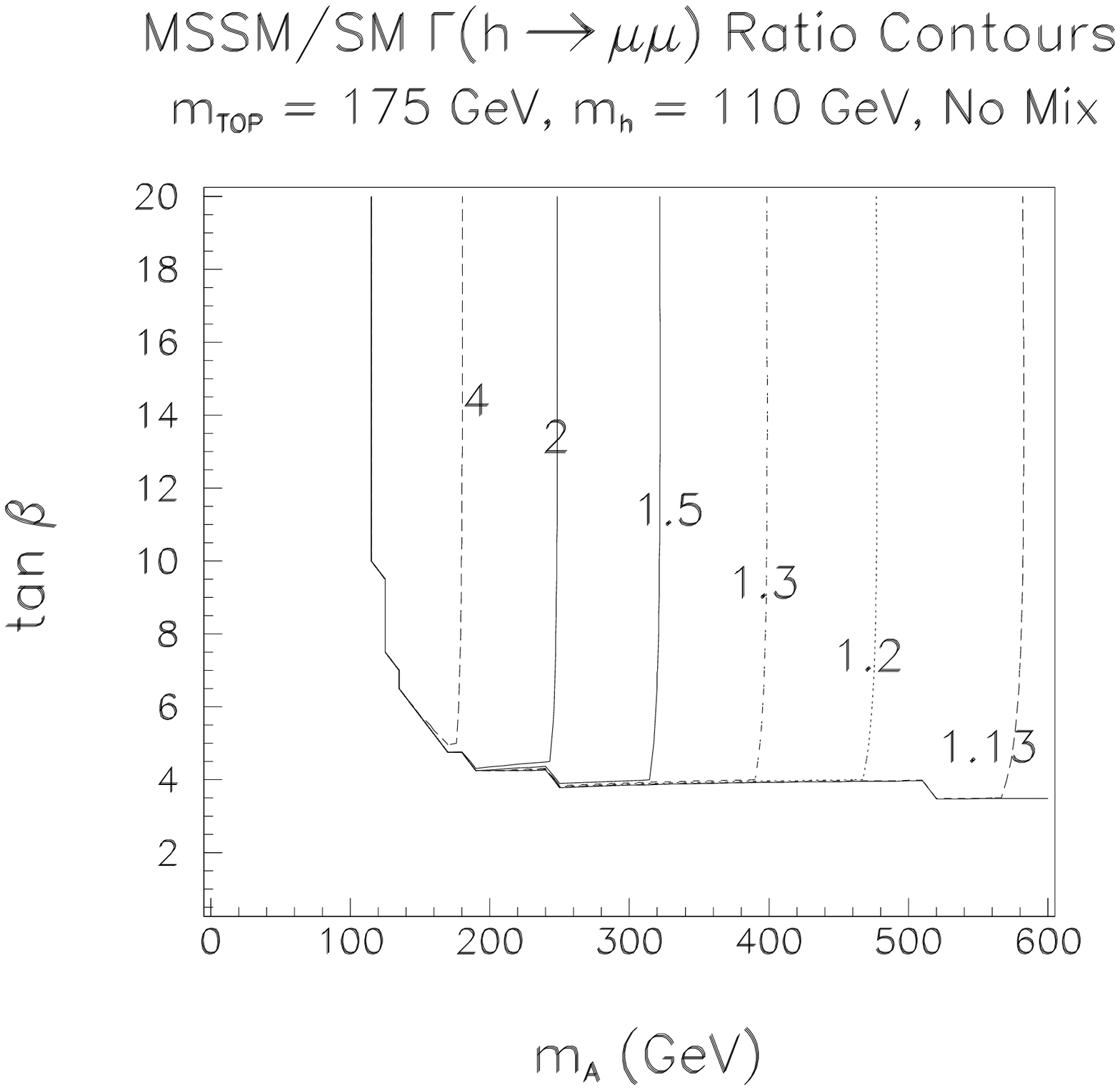,width=3.5in}}
\caption{Constant value contours in $(\mha,\tanb)$ parameter space
for the ratio $\Gamma(\hl\to\mupmum)/\Gamma(\hsm\to\mupmum)$.
We assume ``no mixing'' in the squark sector and present
results for the case of fixed $\mhl=\mhsm=110\gev$. 
For ``maximal mixing'', the vertical contours
are essentially identical --- only the size of the allowed parameter
range is altered. Contours for $\Gamma(\hl\to b\anti b)/\Gamma(\hsm\to b\anti
b)$ are identical.}
\label{mumucontours}
\end{figure}

Finally, let us consider what can be achieved if $e$C 
(or equivalent muon-collider) $\rts=500\gev$
data ($L=200\fbi$), is combined with $s$-channel $\mu$C data.
The most powerful discriminator between the $\hl$ and $\hsm$ turns
out to be $\Gamma(\h\to\mupmum)$.  This is because there
are four statistically independent ratios that
can be used to determine $\Gamma(\h\to\mupmum)$:
\bea
&&{(1)}~~{[\Gamma(\hsm\to\mupmum)\br(\hsm\to
b\anti b)]_{\mu{\rm C}}\over \br(\hsm\to b\anti b)_{e{\rm C}}}\,;\nonumber\\
&&{(2)}~~{[\Gamma(\hsm\to\mupmum)\br(\hsm\to
W\wstar)]_{\mu{\rm C}}\over\br(\hsm\to W\wstar)_{e{\rm C}}}\,;\nonumber\\
&&{(3)}~~{[\Gamma(\hsm\to\mupmum)\br(\hsm\to
Z\zstar)]_{\mu{\rm C}}[\gamhsm]_{e{\rm C}+\mu{\rm C}}
\over\Gamma(\hsm\to Z\zstar)_{e{\rm C}}}\,;\nonumber\\
&&{(4)}~~{[\Gamma(\hsm\to\mupmum)\br(\hsm\to
W\wstar)\gamhsm]_{\mu{\rm C}}\over\Gamma(\hsm\to W\wstar)_{e{\rm
C}}}\,.\nonumber
\eea
At $\mh=110\gev$, the error for $\Gamma(\h\to\mupmum)$ that results
from combining all these determinations is $\sim \pm 4\%$.  (This error
only increases to $\sim\pm 5\%$ if the $\mu$C $s$-channel integrated luminosity
is decreased from $0.4\fbi$ to $0.1\fbi$, for example.)
Given that there is no systematic uncertainty in the prediction
for $\Gamma(\h\to\mupmum)$ ($m_\mu$ is precisely known),
we see from Fig.~\ref{mumucontours} that $\Gamma(\h\to\mupmum)$
provides a $3\sigma$ level of discrimination between the $\hsm$
and the $\hl$ all the way out to $\mha\gsim 600\gev$. Similar results
apply so long a $\mh$ is not in the vicinity of $\mz$. 

The importance of determining the SM-like Higgs properties
with the greatest possible precision argues strongly for having
{\it both} a $e$C and a $\mu$C, especially since very substantial
accumulated luminosity is needed at {\it both} machines
in order to achieve good precision for a model-independent
determination of all the Higgs couplings and its total width.

\bit
\item
What can we learn from the branching ratios of the $\hh,\ha$
in the supersymmetric GUT context?
\eit

Once the $\hh,\ha$ and/or $\hpm$ have been detected (either
in $s$-channel production at a $\mu$C or in pair production
at the $e$C), it will be possible to measure their branching
ratios to various final states with substantial precision.
Refs.~\cite{gk,fengmoroi} have demonstrated that 
such measurements, in combination with
a determination of $\mha$ and one of the gaugino masses,
will allow determination of a large number of the
MSSM parameters, thereby discriminating
between different sets of initial $\mgut$ boundary conditions.
In particular, Ref.~\cite{gk} shows that expected accuracies
for supersymmetric mode decays in pair production
will allow one to discriminate at a very high confidence
level between different boundary condition models.
Thus, Higgs pair production will be a powerful
tool in determining the correct GUT-scale boundary conditions.

Additional issues/questions that we do not have room to address in detail
relate to a Higgs sector with CP-violation.  If the Higgs
sector is found to be CP-violating, then the MSSM is ruled out.
However, the NMSSM and its extensions can have CP violation
in the Higgs sector. Recent work in this area includes:

\bit
\item
The development of a new sum rule for Higgs couplings
and its use to extract limits from LEP-2 for a
CP-violating general (non-MSSM) two Higgs doublet model (2HDM)
or two-doublet plus one singlet (2D1S) Higgs sector \cite{hsumrule}.
The sum rule can be used to show
that LEP-2 data excludes the possibility that there can be
two light Higgs bosons in the CP-violating 2HDM or 
three light Higgs bosons in the 2D1S model.
\item
Development of optimal procedures
for precision measurement of the couplings, including CP-violating
couplings, for a neutral Higgs boson
at a high energy $\epem$ collider by combining $t\anti t \h$ and $Z\h$
production data \cite{cpvcoupnlc}.
One finds that truly excellent errors/limits on all couplings
should be attainable. Prospects for determining the CP properties
of a Higgs boson at the LHC using $t\anti t \h$ associated
production are also not insubstantial \cite{cpvcouplhc}.
\eit

\section{The lightest and next-to-lightest supersymmetric particles}

The phenomenology of low-energy supersymmetry depends crucially on
the properties of the lightest supersymmetric particle (LSP).
In R-parity conserving low-energy supersymmetry, all Standard Model
particles are R-even while their superpartners are R-odd.  Thus,
starting from an initial state involving
ordinary (R-even) particles, it follows that
supersymmetric particles must be
produced in pairs.  In general, these particles are highly unstable
and decay quickly into lighter states.  However, R-parity invariance
also implies that the LSP is absolutely
stable, and must eventually be produced
at the end of a decay chain initiated by the decay of a heavy unstable
supersymmetric particle.

In order to be
consistent with cosmological constraints, a stable LSP is almost
certainly electrically and color neutral. Consequently (with the exception of
a gluino-LSP), the LSP in an R-parity-conserving theory is
weakly-interacting in ordinary matter,
\ie\ it behaves like a stable heavy neutrino
and will escape detectors without being directly observed.
Thus, the canonical signature for conventional R-parity-conserving
supersymmetric theories is missing (transverse)
energy, due to the escape of the LSP.

As noted earlier, in mSUGRA models the LSP is likely to be the $\cnone$
(as opposed to the gravitino) and it
tends to be dominated by its U(1)-gaugino
component (but, see below). A nearly pure
U(1)-gaugino (sometimes called a {\it bino}) will be denoted by $\wt B$.
However, there are some regions of mSUGRA parameter space
where other possibilities for the LSP are realized.
For example, there are regions of mSUGRA parameter space where
the LSP is a chargino.  These regions must be excluded since we reject
the possibility of charged relic particles surviving the early universe.
A sneutrino LSP is possible in mSUGRA models only if $\msnu\lsim
80\gev$. In addition, the requirement that 
relic LSP's do not ``overclose'' the universe by
contributing a mass density that is larger than
the critical density of the universe rules out
additional regions of mSUGRA parameter space.  Requiring 
that the relic density of LSP's constitutes a significant part of the
dark matter would impose further restrictions on the mSUGRA parameter space.

In more general SUGRA
models, the nature of the LSP need not be so constrained.  One can
envision a $\cnone$-LSP which has an arbitrary mixture of gaugino and
higgsino components by relaxing gaugino mass unification.
A nearly pure wino LSP (denoted $\wt W$) is the result if $M_2<M_1$.
A nearly pure higgsino LSP (denoted by $\wt H$) is
possible in the region where the gaugino Majorana masses satisfy
$M_1\simeq M_2\gsim\mu$.  The sneutrino can be a viable LSP
(with no mass restriction as in mSUGRA models),
although it is unlikely to be a major component of the dark matter.
Finally, as described earlier,
the gluino (more strictly speaking an R-hadron containing
the gluino, \eg\ $R^0=\gl g$) can be the LSP \cite{guniondrees2}. 

In simple GMSB models, it was noted earlier that the mass of the gravitino
lies in the eV--keV regime.  Thus, in such models, the
gravitino is the LSP, and the next-to-lightest
supersymmetric particle (NLSP) also plays a crucial role in the phenomenology
of supersymmetric particle production and decay.  Note that unlike the
LSP, the NLSP can be charged.   In GMSB models,
the most likely candidates for
the NLSP are $\cnone$ and $\wt\tau_R^\pm$.~\footnote{In GMSB models, the
$\staur$ tends to be the lightest slepton.  It is lighter than
$\wt e_R$ and $\wt\mu_R$ because of the effect of the larger
$\tau$-lepton-Higgs Yukawa coupling on the
evolution of scalar masses from the messenger scale to the electroweak
scale.  Note that the same Yukawa coupling is responsible for
$\staur$--$\staul$ mixing, so that the light $\stau$ mass eigenstate
would not be a pure $\staur$.  Nevertheless, the $\staur$ component of
the lighter $\stau$ would dominate, so we ignore this distinction in
the notation.}
The NLSP will decay into its superpartner plus a gravitino
[either  $\cnone\to N\gtino$ ($N=\gamma$, $Z$, or $\hl$) or
$\wt\tau_R^\pm\to\tau^\pm\gtino $], with a lifetime that is
quite sensitive to the model parameters.
In a small range of parameter space, it is possible that several of the
next-to-lightest supersymmetric particles are sufficiently degenerate
in mass such that each one behaves as the NLSP.
In this case, these particles will be called co-NLSP's \cite{conlsp}.

Different choices for the identity of the NLSP and its decay rate lead to a
variety of distinctive supersymmetric phenomenologies
\cite{gmsb,eegamgamkane,smartingmsbsearch,lopezgmsb,lepnlcgmsbgam,baerbrhliktata,dicusstau}.
These will be examined in detail later.

A possible exception to the above gravitino LSP scenario is that
discussed earlier. It is possible \cite{raby} to construct a GMSB
model in which the gluino is the LSP and the gravitino is
much heavier (and, thus, phenomenologically irrelevant).
The resulting phenomenology (to be discussed later)
is not dissimilar to that associated with
the SUGRA gluino-LSP case mentioned earlier in association
with the O-II model studied in \cite{guniondrees2}.

Finally, we re-emphasize that restrictions on the SUGAR and GMSB
model parameter spaces can arise from requiring that the relic
LSP density not overclose the universe, but that it is premature
to consider parameter restrictions based upon requiring that
the LSP be the primary component of dark matter; there are
other possible sources of dark matter in all models.

\section{Classes of supersymmetric signals}

Due to a lack of knowledge of the origin and structure of
the supersymmetry-breaking parameters, the predictions of low-energy
supersymmetry depend on a plethora of unknown parameters.
Many details of supersymmetric phenomenology are
strongly dependent on the underlying assumptions of the
model.  Nevertheless, we can broadly classify supersymmetric
signals at future colliders by considering the various theoretical approaches
described earlier. 
In this section, we examine a variety of supersymmetric
signatures, and in the next section we explore their consequences for
experimentation at future colliders.

\subsection{Missing energy signatures}

In R-parity conserving low-energy supersymmetry, supersymmetric particles are
produced in pairs.   The subsequent decay of a heavy supersymmetric particle
generally proceeds via a multi-step decay chain \cite{chain},
ending in the production of at least one
supersymmetric particle which is weakly interacting and escapes the collider
detector.  Thus, supersymmetric particle production yields events that
contain at least two escaping non-interacting particles, leading to a missing
energy signature.~\footnote{We are aware of only two counterexamples. 
First, there are
models in which the $\cnone$ is the LSP and the lightest
neutralino and chargino are nearly degenerate in mass.  If the mass difference
is $\lsim 100$~MeV, then $\cpone$ is long-lived and decays outside the detector
\cite{guniondrees1,guniondrees2}. Second, there is the gluino-LSP
scenario \cite{guniondrees2,raby}, 
the signatures for which will be elaborated upon later.}
(Only the missing transverse energy $\etmiss$ 
can be determined at hadron colliders.)

In conventional SUGRA-based models,
all supersymmetric events contain at least two LSP's which
will escape the collider detector. This leads to the ``smoking-gun''
signature of low-energy supersymmetry: events with large missing
transverse energy
in association with jets and/or leptons. However, there
are two unconventional approaches in which the smoking-gun signature is
absent. First, there is the model in which
the $\cnone$ is the LSP but the lightest
neutralino and chargino are nearly degenerate in mass.  If the mass difference
is $\lsim 100$~MeV, then the 
$\cnone$ is long-lived and decays outside the detector
\cite{guniondrees1,guniondrees2}. Second, there is the gluino-LSP scenario.
Since a hadronic calorimeter measures only the gluino kinetic energy, events in
which the gluino (more precisely, the lightest R-hadron)
is absorbed in or passes through the hadronic calorimeter would also have
substantial missing energy; see later discussion.
However, there would be no jets arising from $\gl$ decays in such models.

In conventional GMSB models, all supersymmetric
events contain at least two NLSP's, and the resulting signature depends on the
NLSP properties.   Four physically distinct possible scenarios emerge:
\begin{itemize}
\item (1)
The NLSP is electrically and color neutral and long-lived, and
decays outside of the detector
to its associated Standard Model partner and the gravitino.
\item (2) The NLSP is the sneutrino and decays invisibly
into $\nu \gtino$ either inside or outside the detector.
\end{itemize}
In either of these two cases, the resulting missing-energy signal is
then similar to that of the SUGRA-based models where $\cnone$ or
$\snu$ is the LSP.
\begin{itemize}
\item (3)
The NLSP is the $\cnone$ and decays inside the detector to $N \gtino$, where
$N=\gamma$, $Z$ or a neutral Higgs boson.
\end{itemize}
In this case, the gravitino-LSP behaves like the neutralino or sneutrino LSP of
the SUGRA models.  But, the missing energy events in this case are
characterized by the associated production of (at least) two $N$'s, one for
each LSP.~\footnote{If the decay of the NLSP is not prompt, it is possible to
produce events in which one NLSP decays inside the detector and one NLSP decays
outside of the detector.}  Note that if $\cnone$ is lighter than the $Z$ (and
$\hl$) then BR$(\cnone\to\gamma \gtino)=100\%$, and all supersymmetric
production will result in missing energy events with at least two associated
photons.
\begin{itemize}
\item  (4)
The NLSP is a charged slepton (typically $\staur$ in GMSB models
if $\mstaur<\mcnone$),
which decays to the corresponding lepton partner and gravitino.
\end{itemize}
If the decay is prompt, then one finds missing energy events with associated
leptons (taus).  If the decay is not prompt, one observes a long-lived
heavy semi-stable charged particle with {\it no} associated missing energy
(prior to the decay of the NLSP).

As noted earlier, there are GMSB 
scenarios in which there are several next-to-lightest
supersymmetric particles (co-NLSP's) that are almost degenerate, in
particular sufficiently degenerate that decays down to the true NLSP
have a very long lifetime.  The canonical example of
this scenario is the case where the right-handed sleptons of the $e$ and
$\mu$ type have mass so close to that of the $\staur$ that they are effectively
stable against chain decays down to the $\staur$. Then any one of the three
types of sleptons can emerge as the end product of a typical supersymmetric
decay chain. The resulting signals can be linear combinations of some of the
above scenarios.  For additional details on the phenomenology of the
co-NLSP's, see Ref.~\cite{conlsp}.

Let us return to the scenario in which
a moderately massive gluino is the LSP. We have noted
that this case can arise in both
SUGRA models \cite{guniondrees2} and in non-conventional GMSB models
\cite{raby}. 
(It is important to note that in the GMSB case, the gravitino is expected
to be much more massive than the gluino and will not enter into
the phenomenology.) Supersymmetric events will terminate with
the production of at least two gluinos that turn into the lightest R-hadron,
most probably the $R^0=\gl g$. 
Since a hadronic calorimeter measures only the R-hadron 
kinetic energy, events in
which the R-hadron is absorbed in the hadronic calorimeter would also have
substantial missing energy that would be an increasingly large
fraction of the total energy as the gluino mass becomes large.
However, since the $\gl$ does not chain decay,
the associated jet production in $\gl\gl$ events would
be very different from that of mSUGRA-based models. Experimental
limits would need to be re-evaluated.

In R-parity violating low-energy SUGRA models, the LSP is unstable.  If the
RPV-couplings are sufficiently weak, then the LSP will decay outside the
detector, and the standard missing energy signal applies.  If the LSP decays
inside the detector, the phenomenology of RPV-models depends on the identity of
the LSP and the branching ratio of possible final state decay products.
If the latter includes a neutrino, then the corresponding
RPV supersymmetric
events would result in missing energy (through neutrino emission) in
association with hadron jets and/or leptons.  However, other decay chains are
possible depending on the relative strengths of $\lambda_L$, $\lambda^\prime_L$
and $\lambda_B$ [see \Eq{rpv}].  Other possibilities include decays into
charged leptons in association with jets (with no neutrinos), and decays into
purely hadronic final states.  Clearly, these latter events would contain little
or no missing energy. If R-parity violation is
present in GMSB models, the RPV decays of the NLSP can easily dominate
over the NLSP decay to the gravitino. In this case, the phenomenology
of the NLSP resembles that of the LSP of SUGRA-based RPV models.

\subsection{Lepton signatures}

Once supersymmetric particles are produced at colliders, they do not
necessarily decay to the LSP (or NLSP) in one step.  The resulting decay chains
can be complex, with a number of steps from the initial decay to the final
state \cite{chain}.
Along the way, decays can produce real or virtual $W$'s, $Z$'s and
sleptons, which then can produce leptons in their subsequent decays.  Thus,
many models yield large numbers of
supersymmetric events characterized by one or more leptons in
association with missing energy and with or without hadronic jets.

One signature of particular note is events containing like-sign
di-leptons \cite{likesign}.
The origin of such events is associated with the Majorana nature of the
gaugino.  For example, $\gl\gl$ production, followed by gluino decay via
\beq
\gl\to q\anti q\cpmone\to
q\anti q\ell^\pm\nu\cnone
\label{gldecay}
\eeq
can result in like-sign leptons since the $\gl$ decay leads with equal
probability to either $\ell^+$ or $\ell^-$.
If the masses and mass differences are both substantial
(which is typical in mSUGRA models, for example), like-sign di-lepton
events will be characterized by fairly energetic jets and isolated
leptons and by large $\etmiss$ from the LSP's.
Other like-sign di-lepton signatures can arise in a similar way from the decay
chains initiated by the heavier neutralinos.

Distinctive tri-lepton signals \cite{bcpttri} can result from
$\cpmone\cntwo\to(\ell^{\pm}\cnone)(Z^*\cnone)$
when $Z^*\to \ell^+\ell^-$.
Such events have little hadronic activity (apart
from initial state radiation of jets off
the annihilating quarks at hadron
colliders).  These events can have a variety of interesting characteristics
depending on the fate of the final state neutralinos.

In GMSB models with a charged slepton NLSP, the decay
$\slep\to\ell\,\gtino$ (if prompt)
yields at least two leptons for every supersymmetric event
in associated with missing energy.
In particular, in models with a $\staur$ NLSP, supersymmetric events
will characteristically contain at least two $\tau$'s.

In RPV models, decays of the LSP
(in SUGRA models) or NLSP (in GMSB models)
mediated by RPV-interactions proportional to $\lambda_L$ and $\lambda^\prime_L$
will also yield supersymmetric events containing charged leptons. However,
if the only significant RPV-interaction is the one proportional to
$\lambda^\prime_L$, then such events would
{\it not} contain missing energy (in contrast to the GMSB signature
described above).

\subsection{$b$-quark signatures}

The phenomenology of gluinos and squarks depends critically on their relative
masses.  If gluinos are heavier, they will decay dominantly into
$q\sq$,~\footnote{In this section, we employ the notation $q\sq$ to mean
either $q\anti{\sq}$ or $\anti{q}\sq$.}
while the squark can decay into quark plus chargino or
neutralino.  If squarks are heavier, the squarks will decay dominantly into
quark plus gluino, while the gluino will decay into the three-body modes
$q\bar q\wt\chi$ (where $\wt\chi$ can be either a neutralino or
chargino, depending
on the charge of the final state quarks).  A number of special cases can arise
when the possible mass splitting among squarks of different flavors is taken
into account.  For example, in mSUGRA models it is often the case
that the third generation squarks are lighter than the squarks of
the first two generations.  If the gluino is lighter than the latter but
heavier than the former, then the only open gluino two-body decay mode
could be $b\sbot$.~\footnote{Although one top-squark mass-eigenstate
($\stopone$) is typically
lighter than $\sbot$ in models, the heavy top-quark mass may result in a
kinematically forbidden gluino decay mode into $t\stopone$.}
In such a case, all $\gl\gl$ events will result in at least
four $b$-quarks in the final state (in associated with the usual missing energy
signal, if appropriate).  
The ${\bf 1}$, ${\bf 24}$ and ${\bf 200}$
Snowmass96 point scenarios of Table~\ref{susymasses} yield precisely
this situation. The $\gl$ decay chains in these three scenarios are:
\begin{eqnarray*}
 {\bf 1:} && ~~\gl \stackrel{90\%}{\to} \sbl\anti b \stackrel{99\%}{\to}
\cntwo b\anti b\stackrel{79\%}{\to} \cnone b\anti
b+(\epem,\mupmum,\nu\anti\nu,q\anti q)\\
 \phantom{{\bf 1:}} && 
\phantom{~~\gl \stackrel{90\%}{\to} \sbl\anti b \stackrel{99\%}{\to}
\cntwo b\anti b}
\stackrel{8\%}{\to} \cnone b\anti b + b\anti b\\
 {\bf 24:} && ~~\gl \stackrel{85\%}{\to} \sbl\anti b
\stackrel{70\%}{\to} \cntwo b\anti b\stackrel{99\%}{\to}
 \hl \cnone b\anti b\stackrel{28\%}{\to} \cnone b\anti b b\anti b \\
\phantom{ {\bf 24:}} && 
\phantom{ ~~\gl \stackrel{85\%}{\to} \sbl\anti b
\stackrel{70\%}{\to} \cntwo b\anti b\stackrel{99\%}{\to}
 \hl \cnone b\anti b}
\stackrel{69\%}{\to} \cnone \cnone\cnone b\anti b \\
 {\bf 200:} && ~~\gl \stackrel{99\%}{\to} \sbl\anti b
\stackrel{100\%}{\to} \cnone b\anti b \\
\label{bchains}
\end{eqnarray*}
More generally, due to the flavor
independence of the strong interactions, one expects three-body
gluino decays into
$b$-quarks in $\sim 20\%$ of all gluino decays.~\footnote{Here we assume the
approximate degeneracy of the first two
generations of squarks, as suggested from the
absence of FCNC decays.  In many models, the $b$-squarks tend to
be of similar mass or lighter than the squarks of the first two generations.}
Additional $b$-quarks can
arise from both top-quark and top-squark decays,
and from neutral Higgs bosons produced somewhere in the
chain decays \cite{bquarkbaer}.

These observations suggest that many supersymmetric events at hadron colliders
will be characterized by $b$-jets in association with missing energy
\cite{snowtheory2,bquarkian}.

\subsection{Signatures involving photons}

In mSUGRA models, most supersymmetric events do not contain isolated
energetic photons.  However, some special regions
of low-energy supersymmetric parameter
space do exist in which final state photons are probable in the decay chains of
supersymmetric particles.  For example, 
if one relaxes the condition of gaugino
mass unification [\Eq{gunif}], and chooses $M_1\simeq M_2$, then the branching
ratio for $\cntwo\to\cnone\gamma$ can be significant \cite{wyler}.
In one particular model of this type
\cite{kane}, the $\cnone$-LSP is dominantly higgsino, while $\cntwo$ is
dominantly gaugino.  Thus, many supersymmetric decay chains end in the
production of $\cntwo$, which then decays to $\cnone\gamma$.  In this picture,
the pair production of supersymmetric particles often
yields two photons plus associated missing energy.  At LEP-2, one can also
produce $\cnone\cntwo$ which would then yield single photon events in
association with large missing energy.

In GMSB models with a $\cnone$-NLSP, all supersymmetric decay chains would end
up with the production of $\cnone$.  
In many models, the branching ratio for $\cnone\to\gamma\gtino$
is significant (and could be as
high as 100\% if other possible two-body decay modes are not kinematically
allowed).  Assuming that $\cnone$ decays inside the
collider detector, supersymmetric pair production would yield
events with two photons in associated with large missing energy.  The
characteristics of these events differ in detail from those of
the corresponding events expected in the model of Ref.~\cite{kane}.

\subsection{Kinks and long-lived sparticles}

In most SUGRA-based models, all heavy sparticles decay promptly
in the decay chain until the LSP is reached.  The LSP is exactly stable and
escapes the collider detector.  However, exceptions are possible.
In particular, if there is a supersymmetric
particle that is just barely heavier
than the LSP, then its (three-body) decay rate
to the LSP will be significantly suppressed and it could be long lived.  For
example, in the models with $M_1> M_2$ \cite{guniondrees1,guniondrees2}
implying $\mcpmone\simeq\mcnone$, the $\cpmone$ can be sufficiently 
long-lived to yield a detectable vertex, or perhaps even exit the detector.

In GMSB models, the NLSP may be long-lived, depending on its mass and
the scale of supersymmetry breaking, $\sqrt{F}$.
The couplings of the NLSP to the
helicity $\pm\half$ components of the gravitino
are fixed by \Eq{gtrel}, as described earlier. 
For $\sqrt F\sim 100$--$10^4\tev$, this coupling is very weak,
implying that all the supersymmetric particles other than the NLSP undergo
chain decays down to the NLSP (the branching ratio for the direct
decay to the gravitino is negligible). The NLSP is unstable and
eventually decays to the gravitino.
For example, in the case of the $\cnone$-NLSP (which is dominated by
its U(1)-gaugino component) one can use \Eq{gtrel}\ to obtain the
decay rate $\Gamma(\cnone\to \gam\gtino)=
m^5_{\tilde\chi_1^0}\cos^2\theta_W/16\pi F^2$. It then follows that
\begin{equation}
(c\tau)_{\cnone\to \gam \gtino}\sim 130
\left({100\gev\over\mcnone}\right)^5
\left({\sqrt F\over 100\tev}\right)^4\mu {\rm m}\,.
\label{ctauform}
\end{equation}
For simplicity, assume that $\cnone\to\gamma\gtino$ is the dominant NLSP
decay mode.  If $\sqrt F\sim 3000\tev$~\footnote{Recall that 
larger values are probably
forbidden in order to avoid overclosing the universe.}
then the decay length for the NLSP is
$c\tau\sim 100$~m for $\mcnone=100\gev$; while
$\sqrt F\sim 100\tev$ implies a short but vertexable decay length.

A similar result is obtained in the case of a charged NLSP.  Thus, if
$\sqrt{F}$ is sufficiently large, the charged NLSP will be semi-stable and
may decay outside of the collider detector.

\subsection{Exotic supersymmetric signatures}

If R-parity is not conserved, supersymmetric phenomenology exhibits
features that are quite distinct from those of the MSSM.
Both $\Delta L\!=\! 1$ and
$\Delta L\!=\! 2$ phenomena are allowed (if L is
violated), leading to neutrino masses
and mixing, neutrinoless double beta decay,
and sneutrino-antisneutrino mixing.
Further, since the distinction between the Higgs and matter
multiplets is lost, R-parity violation permits the mixing of sleptons
and Higgs bosons, the mixing of neutrinos and neutralinos, and the
mixing of charged leptons and charginos, leading to more complicated
mass matrices and mass eigenstates than in the MSSM.

Consequences for collider signatures are numerous.
Most important, the LSP is no
longer stable, which implies that not all supersymmetric decay chains
must yield missing-energy events at colliders.  
All $\cnone$ decays contain visible particles:
\beq
\cnone\to \underbrace{(3j)}_{\lam_B\neq 0}, 
~~\underbrace{(\ell\ell^{(\prime)}\nu)}_{\lam_L\neq 0}, 
~~\underbrace{(\ell 2j,\nu 2j)}_{\lam_L^\prime\neq0}\,.
\label{cnonedecays}
\eeq
Thus, even $\epem\to \cnone\cnone$ pair production becomes visible.
A sneutrino decaying via $\snu\to \nu\cnone$ also becomes visible.
For $\lambda_L\neq 0$, sneutrino resonance production in $\epem$ \cite{schannel}
or $\mupmum$ \cite{fghsnu} collisions becomes possible.
For $\lambda_L^\prime\neq 0$, squarks can be regarded as leptoquarks
since the following processes are allowed:
$e^+\overline u_m\to \overline{\widetilde d}_n\to e^+\overline
u_m$, $\overline\nu\overline d_m$ and
$e^+ d_m \to \widetilde u_n\to e^+d_m$.
(Here, $m$ and $n$ are generation labels, so that $d_2=s$, $d_3=b$,
\etc)
These processes have received much attention during the past year as a
possible explanation for the HERA high-$Q^2$ anomaly \cite{hera}.
The same term responsible for the processes displayed above could also generate
purely hadronic decays for sleptons and sneutrinos:
{\it e.g.}, $\widetilde\ell^-_p\to
\overline u_m d_n$ and $\widetilde\nu_p\to \overline q_m q_n$ ($q=u$ or
$d$).  If such decays were dominant, then the pair production of
sleptons in $e^+e^-$ events would lead to hadronic four-jet
events with jet pairs of equal mass \cite{fourjet}, a
signature quite different from the missing energy signals expected in
the MSSM. Alternatively, $\lambda_L\neq 0$ could result in
substantial branching fractions for
$\slep\to \ell\nu$ and $\snu\to \ell^+\ell^-$ decays.
Sneutrino pair production would then yield
events containing four charged leptons with two lepton pairs of equal mass.

\section{Supersymmetry searches at future colliders}

In this section, we consider the potential for discovering 
and studying low-energy
supersymmetry at future colliders.  Various supersymmetric
signatures have been reviewed earlier, and we now apply these to
supersymmetry searches at future hadron colliders.  
Ultimately, the goal of experimental studies of supersymmetry
is to measure as many of the MSSM-124 parameters (and any additional
parameters that can arise in non-minimal extensions) as possible.  In
practice, a fully general analysis will be difficult, particularly
during the initial supersymmetry discovery phase.  Thus, we focus the
discussion in this section on
the expected phenomenology of supersymmetry at the various
future facilities under a number of different model assumptions.
Eventually, if candidates for supersymmetric phenomena are discovered,
one would plan to utilize precision experimental measurements to map out
the supersymmetric parameter space and uncover the structure
of the underlying supersymmetry-breaking.

\subsection{SUGRA-based models}

We begin with  the phenomenology of mSUGRA.
Of particular importance are the relative sizes of the different
supersymmetric particle masses, which are predicted
in terms of the mSUGRA parameters.
An important generic feature of the resulting superpartner mass spectrum is
that substantial
phase space is available for most decays occurring in a given chain decay of
a heavy supersymmetric particle.

Extensive Monte Carlo studies have determined the region
of mSUGRA parameter space for which direct discovery of
supersymmetric particles at the Tevatron and the LHC
will be possible \cite{baerreview}.
At the hadron colliders, the ultimate supersymmetric mass reach is most
often determined by the searches for the strongly-interacting superpartners
(squarks and gluinos),
although at the Tevatron, the tri-lepton signature will
be stronger in Run-II and at TeV-33 if $\tanb$ is small.
Cascade decays then lead
to events with jets, missing energy, and/or various numbers of leptons.
Gluino and squark masses up to about 400 GeV (or the
equivalent in terms of $\mcpmone,\mcntwo$) can be probed at
the upcoming Tevatron Run-II; further improvements are projected at
the proposed TeV-33 upgrade \cite{tev33msugra}, where supersymmetric
masses up to about 600 GeV can be reached.
The maximum reach at the LHC is attained by searching in
the $1\ell+{\rm jets}+\etmiss$
channel; one will be able to discover squarks and gluinos
with masses up to several TeV \cite{bcptmsugra}. Some particularly important
classes of events include:
\begin{itemize}
\item
$pp\to \gl\gl\to \mbox{jets}+\etmiss$ and 
$\ell^{\pm}\ell^{\pm}$+jets+$\etmiss$,
(the like-sign dilepton signal \cite{likesign}).~\footnote{The possible
predominance of $b$ jets has already been pointed out in \Eq{bchains}.}
For much of parameter
space, the mass difference $\mgl-\mcpmone$ can be determined
from jet spectra end points \cite{likesign,bquarkian,bartlsusy96},
while $\mcpmone-\mcnone$ can be determined from
$\ell$ spectra end points in the like-sign
channel \cite{likesign,bquarkian,bartlsusy96}.
An absolute scale for $\mgl$ can be estimated (within an accuracy of
roughly $\pm 15\%$) by separating
the like-sign events into two hemispheres corresponding to the two $\gl$'s
\cite{likesign}, by a similar separation in the jets+$\etmiss$
channel \cite{bcptmsugra}, or variations thereof \cite{bquarkian,bartlsusy96}.
\item $pp\to \cpmone\cntwo\to(\ell^{\pm}\cnone)(Z^*\cnone)$, which yields
a trilepton + $\etmiss$ final state when $Z^*\to \ell^+\ell^-$;
$\mcntwo-\mcnone$ is easily determined if
enough events are available \cite{bcpttri}.
\item $pp\to\slep\slep\to 2\ell+\etmiss$, detectable
at the LHC for $\mslep\lsim 300\gev$ \cite{bcptmsugra}.
\item Squarks will be pair produced and, for $m_0\gg\mhalf$,  would lead to
$\gl\gl$ events with two extra jets emerging from the
primary $\sq\to q\gl$ decays.
\end{itemize}

The LHC provides significant
opportunities for precision measurements of the mSUGRA parameters
\cite{bquarkian}.  In general, one expects large samples of
supersymmetric events with distinguishing features that allow an
efficient separation from Standard Model backgrounds.  The biggest
challenge in analyzing these events may be in distinguishing one set of
supersymmetric signals from another.  Within the mSUGRA framework, the
parameter space is small enough to permit the untangling of the various
signals and allows one to extract the mSUGRA parameters with some
precision.


Important discovery modes at the $e$C include the following \cite{nlcreport}:
\begin{itemize}
\item $\epem\to \cpone\cmone\to (q\anti q\cnone~\mbox{or}~\ell\nu \cnone)+
(q\anti q\cnone~\mbox{or}~\ell\nu \cnone)$;
\item $\epem\to \slep^+\slep^-\to
(\ell^+\cnone~\mbox{or}~\anti{\nu}\cpone)+
(\ell^-\cnone~\mbox{or}~\nu \cmone)$.
\end{itemize}
In both cases, the masses of the initially produced supersymmetric
particles as well as the final state neutralinos and charginos
will be well-measured.  Here, one is able to make use of the
energy spectra end points and beam energy constraints to make precision
measurements of masses and determine the underlying supersymmetric
parameters.  Polarization of the beams is an essential tool that can be
used to enhance signals while suppressing Standard Model backgrounds.
Moreover, polarization can be employed to separate out various
supersymmetric contributions in order to explore the inherent chiral
structure of the interactions.

The supersymmetric mass reach is limited by the center-of-mass energy of
the $e$C.  For example, if the scalar mass parameter $m_0$ is too large,
squark and slepton pair production will be kinematically forbidden.  To
probe values of $m_0\sim 1$--$1.5\tev$ requires a collider energy in the
range of $\rts> 2$--$3\tev$. It could be that such energies will be
more easily achieved at a future $\mupmum$ collider.

The strength of the lepton colliders lies in their ability to analyze
supersymmetric signals and make precision measurements of observables.
Ideally, one would like to measure the underlying supersymmetric
parameters without prejudice.  One could then test the mSUGRA
assumptions, and study possible deviations.  The most efficient way to
carry out such a program is to first use low machine energies
to study the light supersymmetric
spectrum (lightest charginos and neutralinos and sleptons), and make
model-independent measurements of the associated
underlying supersymmetric parameters.
Once these parameters are ascertained, one can increase
the machine energy and analyze with
more confidence events with heavy supersymmetric particles decaying via
complex decay chains.  Thus, the $e$C and LHC supersymmetric
searches are complementary.

Beyond mSUGRA, the MSSM parameter space becomes more complex.
As discussed briefly earlier, it is possible to 
have non-universality among the scalar mass
parameters (without generating phenomenologically unacceptable FCNC's)
and/or non-universal gaugino masses. We discuss here only 
a few examples of non-universal gaugino mass scenarios.
The first example is the $M_2<M_1$ possibility, for which the
SU(2)-gaugino component is dominant in the lightest chargino and
neutralino \cite{guniondrees1,guniondrees2}.
In this case, the $\cnone$ and $\cpmone$ can be
very degenerate, in which case the visible decay products in the
$\cpone\to \cnone\ldots$ decays will be very soft and difficult to
detect.~\footnote{Of course, in the limit of extreme degeneracy,
the $\cpone$ will be long lived and be detectable
as a visible track in the vertex detector or electromagnetic calorimeter.}
Consequences for chargino and neutralino detection in $\epem$ and
$\mupmum$ collisions, including the importance of the $\epem\to
\gam\cpone\cmone$ production channel, are discussed in
Refs.~\cite{guniondrees1,guniondrees2}.
There is also the possibility that $\mgl\sim \mcpmone\simeq\mcnone$.
The decay products in the $\gl$ decay chain would then be very soft,
and isolation of $\gl\gl$ events would be much more difficult at hadron
colliders than in the usual mSUGRA case. In particular, hard jets
in association with missing energy would be much
rarer, since they would only arise from initial state radiation.
The corresponding reduction in supersymmetric parameter space
coverage for Run-II at the Tevatron and TeV-33 is explored in
Ref.~\cite{guniondrees2}.

As a second example, consider the case where the low-energy gaugino mass
parameters satisfy $M_2\sim M_1$.~\footnote{We remind the reader that
gaugino mass unification at the high-energy scale would predict
$M_2\simeq 2M_1$ [see \Eq{gauginomassrelation}].}  If we also assume
that $\tanb\sim 1$ and $|\mu|<M_1$, $M_2$,~\footnote{To achieve
such a small $\mu$-parameter requires, \eg,  some non-universality among
scalar masses of the form $m_{H_d}^2\neq m_{\sq}^2$, $m_{\slep}^2$.}
then the lightest two neutralinos are nearly a pure photino and
higgsino respectively, \ie,
$\cntwo\simeq\wt\gamma$ and $\cnone\simeq\wt H$.
For this choice of MSSM parameters, one finds that the rate for
the one-loop decay
$\cntwo\to \gam \cnone$ dominates over all tree level decays of $\cntwo$
and $\br(\sel\to e\cntwo)\gg\br(\sel\to e\cnone)$.
Clearly, the resulting phenomenology differs substantially from
mSUGRA expectations.  This scenario was
inspired by the CDF $ee\gam\gam$ event \cite{eegamgamkane}.
Suppose that the $ee\gam\gam$ event resulted from $\sel\sel$ production,
where $\sel\to e\cntwo\to e\gamma\cnone$.
Then one would expect a number of other distinctive supersymmetric
signals to be observable at LEP-2 (running at its maximal energy) and at
Run-II of the Tevatron.  In particular, at 
LEP-2 ($\rts=190\gev$) with $L=500\pbi$
of data one expects more than 50
$\ell\ell+X+\etmiss$ events and at least 3 $\gam\gam+X+\etmiss$
events.  At the Tevatron with $L=100\pbi$ of data, one predicts
more than 30, 2, 15, 4, 2 and 2
events of the types
$\ell\ell+X+\etmiss$, $\gam\gam+X+\etmiss$,
$\ell\gam+X+\etmiss$, $\ell\ell\gam+X+\etmiss$,
$\ell\gam\gam+X+\etmiss$, and $\ell\ell\ell+X+\etmiss$, respectively.
In the above signatures, $X$ stands for
additional leptons, photons, and/or jets.
These signatures can also arise in GMSB models, although the kinematics
of the various events can often be distinguished.

As a final example, we return to the scenario in which the $R^0=\gl g$
is the LSP. What would be the signatures at a hadronic accelerator?
If $\mrzero$ is not large, an initially energetic bare gluino will develop into
a hadronic jet containing an $R^0$. There will be little apparent missing
energy since both the non-$R^0$ components of the jet and the
strongly-interacting $R^0$ itself would deposit energy in the hadronic
calorimeter. Thus, at small $\mrzero$, current limits on $\gl\gl$
production at the Tevatron are no longer
relevant due to the absence of missing energy.
As $\mrzero$ increases, the $R^0$ takes a larger
and larger fraction of the gluino's initial energy and the
energy of the $R^0$ as a fraction of $\mrzero$ that appears in
the calorimeter becomes smaller and smaller. The maximum energy that
can be deposited is the full kinetic energy, 
$KE=\mrzero [1/\sqrt{1-\beta^2}-1]$. For full deposit of all the KE, the gluino
would have to be effectively stopped in the detector. The rest mass energy
would then effectively disappear, leading to substantial missing
energy. However, from our earlier discussions, it must be remembered that the 
path length $l_p$ for the $R^0$ becomes large at large $\mrzero$,
so that an $R^0$ with $\mrzero\gsim 1\tev$ is likely to
simply pass through the detector leaving only a small amount of
deposited energy, implying even larger missing energy.
Whichever is the case, 
the Tevatron limits will again fail to apply, not because of an
absence of missing energy, but rather
because of the small amount of visible energy
associated with the gluino-jet --- the only visibly energetic jets
would be those associated with initial state radiation.
For intermediate $\beta$ values, one could search for delayed energy
deposits in the hadronic calorimeter (with an unusual profile
associated with the fact that the $R^0$ would tend to deposit most
of its KE at the stopping point).  In either case, one could
envision searching for the $R^0$ using 
a high luminosity, high energy $\rts\gsim$ several TeV beam dump experiment, 
in which mass spectrometer
or chemical analysis of the dump could be used to search for the $R^0$-nuclear
isotopes that would be formed. However, if $\mrzero$ is above 100 GeV
or so, the predicted penetration or stopping 
length (as discussed earlier) would imply the need for a very large beam dump.

Of course, if squarks are not too heavy, $\sq\sq$
production will have substantial rate (if not at the Tevatron, then
at the LHC) and each $\sq$ will decay to $q\gl$.  A heavy $\gl$-LSP
will yield missing energy as described above and if $\msq-\mgl$
is substantial there will also be significant associated jet activity.
Tevatron limits employing the jets plus $\etmiss$ signal would presumably
apply.

\subsection{GMSB-based models}

The collider signals for GMSB models depend critically
on the NLSP identity and its lifetime (or equivalently, its decay
length [which depends on the value of $\sqrt{F}$ as shown in
\Eq{ctauform}]).   In this regard,
it should be recalled that values of $\sqrt F$ as large as $10^3-10^4\tev$
are certainly not disfavored in the GMSB context.
Such large values of $\sqrt F$ imply very long decay lengths for the NLSP.
Thus, we examine the phenomenology of 
both promptly-decaying and longer-lived NLSP's. 
In the latter case,
the number of decays where one or both NLSP's decay within a radial
distance $R$ is proportional to $[1-\exp(-2R/c\tau)]\simeq 2R/(c\tau)$.
For large $c\tau$, most decays would be non-prompt, with many
occurring in the outer parts of the detector or completely outside the
detector. To maximize sensitivity to GMSB models
and fully cover the $(\sqrt F,\Lambda)$ parameter space,
we must develop signals sensitive to decays that are delayed,
but not necessarily so delayed as to be beyond current detector
coverage and/or specialized extensions of current detectors.

In the discussion below,
we focus on various cases, where the NLSP is a neutralino dominated by
its U(1)-gaugino (bino) or Higgsino components, and where the NLSP is
the lightest charged slepton (usually the $\staur$).  We first address
the case of prompt decays, and then indicate the appropriate
strategies for the case of a longer-lived NLSP.

\begin{itemize}
\item {{\bf Promptly-decaying NLSP:} $\bold{\cnone\simeq\wtil B}$}
\end{itemize}

We focus on the production of the neutralinos, charginos, and sleptons
since these are the lightest of the supersymmetric
particles in the GMSB models.  The possible decays of the NLSP in this
case are: $\wt B\to \gamma\gtino$ or $\wt B\to Z\gtino$.  The
latter is only relevant for the case of a heavier NLSP, and 
is, in any case, suppressed by a factor of $\tan^2\theta_W$; it will be
ignored in the following discussion.

At hadronic colliders,
the $\cnone\cnone$ production rate is small, but rates for
$\cpone\cmone\to W^{(\star)}W^{(\star)}\cnone\cnone\to W^{(\star)}W^{(\star)}
\gam\gam+\etmiss$,
$\slepr\slepr\to \ell^+\ell^-\cnone\cnone\to\ell^+\ell^-\gam\gam+\etmiss$,
$\slepl\slepl\to\ell^+\ell^-\cnone\cnone\to \ell^+\ell^-\gam\gam+\etmiss$,
\etc\ will all be substantial.  Implications for GMSB phenomenology at
the Tevatron have been studied in
Refs.~\cite{gmsb,baerbrhliktata,smartingmsbsearch,chenguniongampub,chenguniongamprelim}.
As noted earlier, it is possible to envision GMSB parameters such that
the $ee\gam\gam+\etmiss$ CDF event corresponds to slepton pair
production followed by $\cnone\to\gamma\gtino$
 \cite{gmsb,eegamgamkane,smartingmsbsearch,lopezgmsb}.
However, in this region of GMSB parameter space, other supersymmetric
signals should be prevalent, such as
$\snul\slepl\to\ell\gam\gam+\etmiss$ and
$\snul\snul\to\gam\gam+\etmiss$.
The $\cntwo\cpmone$
and $\cpone\cmone$ rates would also be significant and lead to
$X\gam\gam+\etmiss$ with
$X=\ell^\pm,\ell^+\ell^{\prime\,-},\ell^+\ell^-\ell^{\prime\,\pm}$.
Limits on these event rates from current CDF
and D0 data already eliminate much of the parameter space that could lead
to the CDF $ee\gam\gam$ event \cite{limitsoneegg}.

At LEP-2/$e$C, the rate for the simplest signal,
$\epem\to \cnone \gtino \to \gam+\etmiss$, is expected to be very
small  \cite{lepnlcgmsbgam}.  A more robust
channel is $\epem\to \cnone\cnone\to \gam\gam+\etmiss$ with
a (flat) spectrum of photon energies in the range ${1\over
4}\rts(1-\beta)\leq E_\gam\leq {1\over 4}\rts(1+\beta)$.
Ref.~\cite{kanelep} points out that there might be a
small excess of such events in the current LEP-2 data,
but the level of background requires careful evaluation
\cite{mrennagmsb}.

\begin{itemize}
\item {{\bf Promptly decaying NLSP:} $\bold{\cnone\simeq\wtil H}$}
\end{itemize}

The possible decays of the NLSP in this case are: $\wtil H\to \gtino
+\hl,\hh,\ha$, depending on the Higgs masses.  If the corresponding
two-body decays are not kinematically possible, then
three-body decays (where the corresponding Higgs state is virtual) may
become relevant.  However, in realistic cases, one expects $\cnone$ to
contain small but non-negligible gaugino components, in which case
$\cnone\to\gtino\gamma$ would dominate all three-body decays.
In what follows, we assume that the two-body decay $\wtil
H\to\gtino\hl$ is kinematically allowed and dominant.
The supersymmetric signals that would emerge
at both Tevatron/LHC and LEP-2/$e$C would then be
$4b+ X + \etmiss$ final states, where $X$ represents the
decay products emerging from the cascade chain decays
of the more massive superparticles. Of course, at LEP-2/$e$C direct
production of higgsino pairs, $e^+e^-\to \wt H\wt H$ (via virtual
$s$-channel $Z$-exchange) would be possible in general, leading to
pure $4b+\etmiss$ final states.

\begin{itemize}
\item {{\bf Promptly decaying NLSP:} $\bold{\slepr}$}
\end{itemize}

The dominant slepton decay modes
are: $\slepr^\pm\to \ell^\pm \gtino$ and
$\slepl^\pm\to\ell^\pm \cnonestar\to
\ell^\pm(\slepr^\pm\ell^\mp)^\prime
\to \ell^\pm(\ell^\pm\ell^\mp)^\prime \gtino$.
The $\cnone$ will first decay to $\ell \slepl$ and $\ell\slepr$, followed
by the above decays.

At both the Tevatron/LHC and LEP-2/$e$C typical pair production events
will end with
$\slepr\slepr\to \ell^+\ell^-+\etmiss$, generally in association with a variety
of cascade chain decay products. The lepton energy spectrum
will be flat in the $\slepr\slepr$ center of mass.
Of course, pure $\slepr\slepr$ production is possible
at LEP/$e$C and the $\slepr\slepr$ center of mass would be
the same as the $\epem$ center of mass. Other simple signals at LEP/$e$C
would include $\slepl\slepl\to 6\ell +\etmiss$.

As noted earlier, if a slepton is the NLSP, it is most likely to
be the $\staur$.  If this state is
sufficiently lighter than the $\wt e_R$ and $\wt \mu_R$,
then $\wt e_R \to e \staur\tau$ and
$\wtil \mu_R\to \mu \staur\tau$ decays
(via the $\wtil B$ component of the mediating
virtual neutralino) might dominate over
the direct $\wt e_R\to e\gtino$ and $\wt \mu_R\to \mu\gtino$ decays
and all final states would cascade to $\tau$'s. The relative
importance of these different possible decays has been examined
in Ref.~\cite{smartinstau}. A study of this scenario at LEP-2 has
been performed in Ref.~\cite{dicusstau}.

\begin{itemize}
\item {{\bf Longer-lived NLSP:} $\bold{\slepr}$}
\end{itemize}

If the $\slepr$ decays mainly
before reaching the electromagnetic calorimeter,
then one should look for a charged lepton
that suddenly appears a finite distance from the interaction region,
with non-zero impact parameter as measured by either the vertex detector
or the electromagnetic calorimeter. Leading up to this decay would be a heavily
ionizing track with $\beta<1$ (as could be measured if a magnetic
field is present).

If the $\slepr$ reaches the electromagnetic
and hadronic calorimeters, then it
behaves much like a heavy muon, presumably
interacting in the muon chambers or
exiting the detector if it does not decay first.
Limits on such objects should be pursued.  
Limits are currently
quoted by D0 and CDF for pseudo-stable charged strongly produced
particles (stable quarks, squarks, . .. ), but the rates
predicted in GMSB models for $\slepr$ pairs are very different.
There will be many sources of $\slepr$ production, including
direct slepton pair production, and cascade decays resulting from the
production of gluinos, squarks, and charginos.
A detailed calculation is in progress \cite{chengunionslep}.
Based on the slepton pair cross section at the Tevatron, we estimate that
a charged pseudo-stable $\slepr$
can be ruled out with a mass up to about $80$--$100\gev$.  Similar limits
can probably be extracted from LEP-2 data. Including the other
production mechanisms should yield a significant extension of the excluded
mass range in many GMSB models.

\begin{itemize}
\item {{\bf Longer-lived NLSP:} $\bold{\cnone}$}
\end{itemize}

This is a much more difficult case. As before, we assume that the
dominant decay of the NLSP in this case is $\cnone\to\gamma\gtino$.
Clearly, the  sensitivity of detectors to
delayed $\gam$ appearance signals will be of great importance.
As noted earlier, if $\sqrt F$ is so large that
the $\cnone$ escapes the detector before
decaying, then the corresponding missing energy signatures are the
same as those occurring in SUGRA-based models.  

\begin{figure}[h]
\centering
\mbox{\epsfig{file=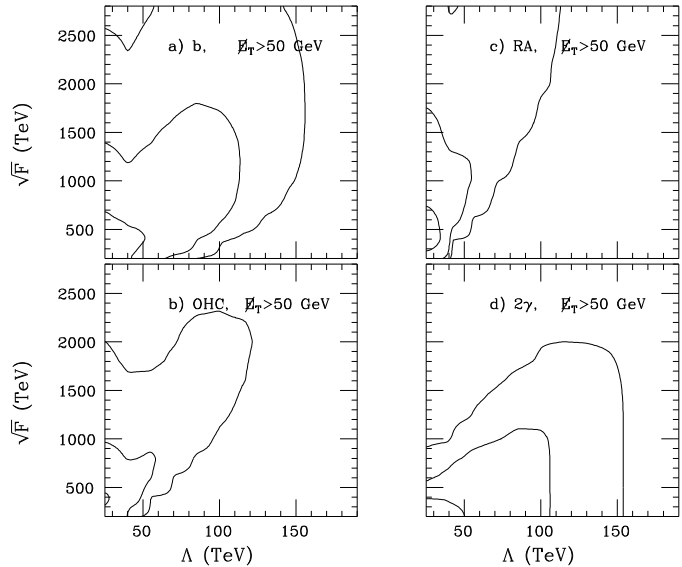,width=3.5in}}
\caption{
We present cross section contours in the $(\protect\sqrt F,\Lambda)$
parameter space for the (a) impact-parameter ($b$), (b) outer-hadronic (OHC),
(c) roof-array (RA), and (d) prompt-two-photon ($2\gam$) signals.
Contours are given at $\sigma=0.16\fb$ (innermost), 2.5~fb (middle), and $50\fb$
(outermost).
Thus, the middle (outermost) contour corresponds to 5 events at
Run-II/$L=2\fbi$ (Tev-33/$L=30\fbi$).
In this figure we have taken $\etmin=50\gev$ for all signals.}
\label{delayfig}
\end{figure}

Consider first what is possible at the Tevatron 
without making use of the delayed $\gam$ appearance signals.
First, for any magnitude of $\sqrt F$, 
the jets plus missing energy signal at the
Tevatron will only be viable in the limited region of
GMSB model parameter space \cite{chenguniongampub} characterized by
the GMSB $\Lambda$ parameter being $\lsim 30\tev$ ($\lsim 40\tev$)
in Run-II (at Tev-33); the standard
tri-lepton signal will be viable for $\Lambda\lsim 65\tev$ ($75\tev$)
\cite{chenguniongamprelim}. The prompt-two-photon plus missing
energy signal deriving from the many sources of
prompt $\cnone\to\gam\gtino$ decays discussed earlier 
is also only visible in a limited portion of
parameter space. The regions of viability for Run-II and at Tev-33
are illustrated in Fig.~\ref{delayfig}.
Thus, even
after combining the jets-plus-missing-energy, tri-lepton and prompt-two-photon
signals, there will be a 
significant region of $(\sqrt{F},\Lambda)$ parameter space
that cannot be probed without using the delayed $\cnone\to \gam\gtino$
decays \cite{chenguniongampub,chenguniongamprelim}.

Thus, it is important that the collider detectors develop signals that
are sensitive to the delayed $\cnone\to \gam\gtino$ decays.
The ability to search for delayed-decay signals
is rather critically dependent upon the detector 
design. The possible signals \cite{chenguniongampub,chenguniongamprelim}
include the following: (a) searching for events where the delayed-decay
photon is identified by a large (transverse)
impact parameter as it passes into the electromagnetic calorimeter;
(b) looking for isolated energy deposits
(due to the $\gam$ from a delayed $\cnone$ decay) 
in outer hadronic calorimeter cells, as possible with the D0 detector;
and (c) looking for delayed decays where the $\cnone$ 
decay occurs outside the main detector and the photon from the decay
is observed in a scintillator (or similar
device) array place at a substantial distance from the detector.
The observed signal will always contain missing energy from the emitted
$\gtino$ or un-decayed $\cnone$, 
which will be especially useful in reducing potential
backgrounds. In all cases, timing information for the detected photon
would be extremely valuable.
Only if these (or closely related) techniques are employed can one
detect GMSB supersymmetry in the bulk (and possibly preferred portion)
of parameter space where neither $\sqrt F$ nor $\Lambda$ is small.

A first estimate of the utility of the above signals,
is given in Fig.~\ref{delayfig}, taken from Ref.~\cite{chenguniongamprelim}.
In this figure we display cross section contours in the minimal GMSB model
for the D0 detector for the three types
of signals described above. Cuts have been imposed that should eliminate
backgrounds. (A D0 collaboration study is needed for confirmation.)
The cuts require a certain number of jets $n_j$
(defined with $\Delta R_{\rm coal}=0.5$ and
required to have $p_T\geq 25\gev$ and $|\eta|\leq 3.5$) and 
one or more delayed
photons (required to have $E_\gam\geq 15\gev$ and be separated
from jets by $\Delta R\geq 0.5$) and a minimum
amount of missing transverse energy $\etmiss$. For all three delayed-decay
signals (a), (b) and (c) presented in Fig.~\ref{delayfig},
we require $n_j\geq 3$ and $\etmiss\geq 50\gev$. 

(a) The impact parameter ($b$) signal is defined by requiring
that there be one or more $\cnone\to\gam\gtino$ decays in which
the impact parameter $b$ of the photon (as measured using the pre-shower
and electromagnetic calorimeter) is $\geq 2$~cm.~\footnote{Since
the expected resolution in $b$ is $\sim 0.2$~cm, this corresponds
to a 10$\sigma$ excursion from the primary interaction point.}

(b) The outer hadronic calorimeter (OHC) signal is defined by requiring
that there be one or more $\cnone\to \gam \gtino$ decays in central
OHC cells.

(c) The roof array (RA) signal is defined by requiring that
there be one or more $\cnone\to\gam\gtino$ decays in which the
photon emerges at a vertical height above 6.5 m (\ie\ outside the detector,
in particular, beyond the muon chambers)
and passes through a 38 m $\times$ 28 m rectangular detection array 
centered vertically above the interaction point at a distance
of 16.5 m (corresponding to the height of the roof of the D0 detector hall).

Assuming that the backgrounds to all these signals are negligible after
our very strong cuts, 5 events will be adequate
to claim discovery.  Fig.~\ref{delayfig} shows that in Run-II (the middle
contours)
the impact parameter signal will cover almost all of the $\sqrt F\lsim 1500\tev$
and $\Lambda\lsim 100\tev$ portion of parameter space.
The RA and OHC signals would provide important backup and confirmation
for the impact parameter signal when $\Lambda\lsim 50\tev$. At Tev-33, the
regions for which the $b$, OHC and RA signals are viable expand greatly.
In combination, they provide sensitivity to GMSB supersymmetry throughout the
preferred $\sqrt F\lsim 3000\tev$, $\Lambda\lsim 150\tev$ parameter region
for the model examined. Clearly it is
of great importance for the D0 and CDF experiments to examine these
signals more closely and develop reliable background estimates. 

Limits on delayed-decay photon signals from Run-I data ($L\sim 100\pbi$) require
analysis of the OHC signal.  A study is underway \cite{mani}.
The roof array was, of course, not present,
and the impact parameter resolution was extremely poor due to the absence
of a preshower.  

\begin{itemize}
\item {{\bf Gluino-LSP}}
\end{itemize}
The phenomenology for this variant \cite{raby} of GMSB model would 
be much as described in the case of a SUGRA model with a gluino-LSP.
The primary distinctions would arise from differences in the
detailed mass spectra for the heavier sparticles.

\subsection{RPV models}

In R-parity violating models, the LSP is no longer stable.~\footnote{We
assume that the gravitino is not relevant for phenomenology, as in
SUGRA-based models.}
The relevant signals depend upon the nature of the LSP decay.  The
phenomenology depends on which R-parity violating couplings [\Eq{rpv}]
are present. For our discussion here, we assume that it is the $\cnone$
that is the LSP, although this is no longer required (by cosmology, \etc)
if the LSP is not stable.

\begin{itemize}
\item At the Tevatron and LHC \cite{likesignrviol}:
\end{itemize}

If $\lambda_B\neq 0$, then $\cnone\to 3j$ (where $j=$jet). 
The large jet backgrounds
imply that we would need to rely on the like-sign dilepton signal
(which would still be viable despite the absence of missing energy
in the events).  In general,
this signal turns out to be sufficient for supersymmetry discovery
out to gluino masses somewhat above 1 TeV.
However, if the leptons of the like-sign dilepton signal are very soft, 
then the discovery reach would be much reduced.~\footnote{Soft leptons
would occur in models where $\mcpmone\sim\mcnone$ (which requires
non-universal gaugino masses \cite{guniondrees1,snowtheory2}).}
This is one of the few cases where one could miss discovering low-energy
supersymmetry at the LHC. 
If $\lambda_L$ dominates $\cnone$ decays,
$\cnone\to \mu^\pm e^\mp\nu,e^\pm e^\mp \nu$, and there would be many
very distinctive multi-lepton signals.
If $\lambda_L^\prime$ is dominant, then
$\cnone\to \ell jj$ and again there would
be distinctive multi-lepton signals.

More generally, many normally invisible signals become visible.
An important example is sneutrino pair production.
Even if the dominant decay of the sneutrino is
$\snu\to\nu\cnone$ (which is likely if $\msnu>\mcnone$),
a visible signal emerges from the $\cnone$ decay as sketched above.
Of course, for large enough $\lambda_L$ or $\lambda_L^\prime$
the $\snu$'s would have significant branching ratio for
decay to lepton pairs or jet pairs, respectively.
Indeed, such decays might dominate if $\msnu<\mcnone$.

\begin{itemize}
\item At LEP-2, an $\epem$ collider or a $\mu^+\mu^-$ collider:
\end{itemize}

The simplest process at the lepton colliders is:
\beq
\epem\to\cnone\cnone\to \underbrace{(3j)(3j)}_{\lambda_B},
~~\underbrace{(2\ell\nu)(2\ell\nu)}_{\lambda_L},
~~\underbrace{(\ell jj)(\ell jj)}_{\lambda_L^\prime}
\label{simppro}
\eeq
(or the $\mupmum$ collision analogue),
where the relevant RPV coupling is indicated below the corresponding signal.
Substantial rates for
equally distinctive signals from production of more massive 
supersymmetric particles (including sneutrino pair production)
would also be present.  All these processes (if kinematically allowed)
should yield observable supersymmetric signals.  Some limits from LEP data 
already exist \cite{lepnlcrpv}.

Of particular potential importance for non-zero $\lambda_L$ is
\hbox{$s$-channel}
resonant production of a sneutrino in $e^+e^-$ \cite{schannel}
and $\mupmum$ \cite{fghsnu} collisions.
At a muon collider, this process is detectable down
to quite small values of the appropriate $\lambda_L$
(due to the very excellent beam energy resolution possible
at a muon collider), and could be
of great importance as a means of actually determining the
R-parity-violating couplings. Indeed, for small R-parity-violating
couplings, absolute measurements of the couplings
through other processes are extremely difficult. This
is because such a measurement would typically require the R-parity-violating
effects to be competitive with an R-parity-conserving process of known
interaction strength. (For example, R-parity-violating
neutralino branching ratios constrain only ratios of the R-parity-violating
couplings.) Since sneutrino pair production would have been observed
at the LHC, an $\epem$ collider and/or the $\mu^+\mu^-$ collider, 
it would be easy to center on
the sneutrino resonance in order to do the crucial sneutrino factory
measurements.

\begin{itemize}
\item At HERA:
\end{itemize}

Squark production via R-parity violating
couplings  could be an explanation for the HERA anomaly
at high $x$ and large $Q^2$ \cite{hera}.
For example, if $(\lambda_L^\prime)_{113}\sim 0.04$--$0.1$,
a leptoquark signal from
$e^+ d_R\to \wtil t_L\to e^+ d$ would be detected if $m_{\wtil t_L}\lsim
220\gev$. 

\bit
\item Delayed RPV decays:
\eit

If the RPV coupling responsible for the decay of the LSP is sufficiently
small, but not too small, the LSP decay might not be prompt but could
still occur inside the detector or at least not far outside the detector.
In this case, the same techniques discussed in the GMSB
models for observing delayed $\cnone\to\gam\gtino$ decays could prove
very valuable. Delayed $\cnone$ decay OHC energy deposits might occur, 
leptons~\footnote{If the RPV interaction leads to $\cnone\to 3j$,
then the ability to measure the impact parameters of jets would
be important, if possible with sufficient accuracy.}
from a delayed $\cnone$ decay 
might be observed (using the electromagnetic calorimeter and its preshower)
to have non-zero impact parameter, and the $\cnone$ could decay outside
the detector but before reaching the roof array. Especially
at lower energy machines such as the Tevatron, a large extension in
parameter space coverage would result from implementing these non-standard
detection modes. Further, the relative abundance of events obtained using
these different techniques, especially if supplemented with timing
information, could allow a determination of the RPV coupling strength.
In many scenarios, such determination would not be possible by any other means.

Finally, we note that 
one should not rule out the possibility that RPV couplings
could be present in GMSB models.  Unless, the RPV couplings are rather
small, the NLSP will decay via the RPV decay channels with
a much higher rate than to channels containing the $\gtino$ \cite{cpwnew},
and all the RPV phenomenology described earlier will apply. However,
it is also possible that RPV and $\gtino$ decays of the NLSP
will take place at competitive rates. Delayed-decay signals could prove
crucial to a full understanding of such a situation.

\section{Summary and Conclusions}

Low-energy supersymmetry remains the most elegant solution to the
naturalness and hierarchy problems, while
providing a possible link to Planck scale physics and the unification of
particle physics and gravity.  Nevertheless, the
origin of the soft supersymmetry breaking
terms and the details of their structure remain a mystery.
There are many theoretical ideas, but we still cannot be certain which
region of the MSSM-124 parameter space (or some non-minimal extension
thereof) is the one favored by nature.
The key theoretical breakthroughs will surely
require experimental guidance and input.

Thus, we must rely on our experimental colleagues to uncover evidence
for supersymmetry at present or future colliders.
Canonical supersymmetric signatures at
future colliders are well analyzed and understood.  Much of the recent
efforts, described in the previous section, have been directed at trying to
develop strategies for
precision measurements to prove the underlying supersymmetric structure
of the interactions and to distinguish among models.  However, we
are far from understanding all possible facets of MSSM-124 parameter
space (even restricted to those regions that are phenomenologically
viable).  Moreover, the phenomenology of non-minimal and alternative
low-energy supersymmetric models (such as models with R-parity
violation) and its consequences for collider physics have
only recently begun to attract significant attention.
The variety of possible non-minimal models of low-energy supersymmetry
presents an additional challenge to experimenters who plan on searching
for supersymmetry at future colliders.

If supersymmetry is discovered, it will provide a gold mine of
experimental signals and theoretical analyses.
The many phenomenological manifestations
and parameters of supersymmetry suggest that many years of experimental
work will be required before it will be possible to determine
the precise nature of supersymmetry-breaking and its
implications for a more fundamental theory of particle interactions.

\section*{Acknowledgements}

This work was supported in part by the Department of Energy and
by the Davis Institute for High Energy Physics. I would like
to acknowledge the important contributions resulting
from collaboration with H.E. Haber on Ref.~\cite{gunhabper}.

\end{document}